\documentstyle[preprint,aps,epsfig]{revtex}

\tightenlines
 
\begin{document}

\title{Pion and sigma meson properties in a relativistic quark model}

\author{Amand Faessler \footnotemark[1], 
Th. \ Gutsche \footnotemark[1],  
M. \ A. \ Ivanov \footnotemark[2], 
V. \ E. \ Lyubovitskij \footnotemark[1], 
P. \ Wang \footnotemark[1]
\vspace*{0.4\baselineskip}}
\address{
\footnotemark[1]
Institut f\"ur Theoretische Physik, Universit\"at T\"ubingen, \\
Auf der Morgenstelle 14,  D-72076 T\"ubingen, Germany 
\vspace*{0.3\baselineskip}\\ 
\footnotemark[2]
Bogoliubov Laboratory of Theoretical Physics, \\ 
Joint Institute for Nuclear Research, 141980 Dubna, Russia
\vspace*{0.3\baselineskip}\\} 

\maketitle
 
\vskip.5cm

\begin{abstract} 
A variety of strong and electroweak interaction properties of the pion
and the light scalar $\sigma$ meson are computed in a relativistic
quark model.
Under the assumption that the resulting
coupling of these mesons to the constituent quarks is identical,
the $\sigma$ meson mass is determined as $M_\sigma=385.4$ MeV. 
We discuss in detail the gauging of the non-local
meson-quark interaction and calculate the electromagnetic form factor
of the pion and the form factors of the $\pi^0\to\gamma\gamma$
and $\sigma\to\gamma\gamma$ processes. We obtain explicit
expressions for the relevant form factors  and evaluate
the leading and next-to-leading orders for large Euclidean  
photon virtualities. Turning to the decay properties of the $\sigma$
we determine the width of the electromagnetic
$\sigma\to\gamma\gamma$ transition and discuss the strong decay
$\sigma\to\pi\pi$. In a final step we compute the nonleptonic decays
$D\to\sigma\pi$ and $B\to\sigma\pi$ relevant for the possible observation  
of the $\sigma $ meson. All our results are compared to available 
experimental data and to results of other theoretical studies.
\end{abstract} 

\vskip.5cm


\vskip.5cm


\section{Introduction}

During the last few years the physics of isoscalar scalar mesons and
in particular of the $\sigma$ meson has
received an revival of interest due to substantial progress in experimental 
and theoretical activities~\cite{PDG} (for a status report see, for example, 
Ref.~\cite{Sigma_review}).
Originally, the light scalar meson 
$\sigma$ was introduced as the chiral partner of the pion in the
two-flavor linear $\sigma$-model~\cite{Schwinger,Gell-Mann_Levy}. 
The linear $\sigma$-model fulfils the chiral commutation relations, contains
the partial conservation of the axial current (PCAC) and has a manifestly 
renormalizable Lagrangian. 
In approaches based on the linear realization of chiral symmetry (see, 
for example,~\cite{Nambu,Schechter}) the $\sigma$ meson serves to 
create spontaneous breaking of chiral symmetry, it generates
the constituent quark mass and it is a relevant ingredient in explaining
low-energy phenomenology 
($\pi\pi$ scattering, $\Delta I=1/2$ enhancement in 
$K\rightarrow \pi\pi$, attractive interaction between 
baryons in nuclear matter, etc.).
On the other hand, the use of the linear or 
non-derivative $\sigma$-model Lagrangian leads to well-known 
difficulties.
For example, current-algebra 
results cannot be reproduced at the tree level and can only be generated
by mixing up different orders in the chiral expansion. 
For this reason, it was suggested~\cite{Weinberg} that the  
linear $\sigma$-model Lagrangian is modified in such a fashion that the 
results of current algebra are already produced at the tree level, while
at the same time a clear chiral power counting is obtained.
This modification is based on a
unitary, chiral field-dependent transformation of the quark/nucleon 
field, which eliminates the non-derivative linear coupling of $\pi$ and 
$\sigma$ and replaces it by a nonlinear derivative coupling of 
the chiral rotation vector, identified as the new pion field.
This construction also serves as a basis for the formulation of
chiral perturbation theory (ChPT)~\cite{ChPT}, which is now considered as 
the realistic effective theory of low-energy hadron 
dynamics. In the context of the nonlinear realization of chiral symmetry
a light $\sigma$-meson might be treated as a resonance
in the $\pi\pi$-system~\cite{Colangelo,Oller}. Alternatively, for the linear 
case the $\sigma$ can either be described as a member of a 4-quark 
multiplet~\cite{Sigma_review} or as quark-antiquark resonance~\cite{Kleefeld}.
The different mechanisms for generating a light $\sigma $ do not necessarily
exclude each other, but could in turn be related in a way
which is not completely understood yet.

Recently, the E791 Collaboration at Fermilab~\cite{E791} and the BES 
Collaboration at BEPC~\cite{BES} reported on evidence for a light and 
broad scalar resonance in nonleptonic cascade decays of heavy mesons. 
In the Fermilab experiment it was found that the $\sigma$ meson is rather
important in the $D$ meson decay $D\rightarrow 3\pi$~\cite{E791}.
In a coherent amplitude analysis of the $3\pi$ Dalitz plot the scalar
resonance is determined with $478_{-23}^{+24}\pm17$ MeV and total width 
$324_{-40}^{+42}\pm 21$ MeV. A fraction $f=(46 \pm 11)\%$ of the decay mode 
$D^+\rightarrow\pi^+\pi^-\pi^+$ is generated by the intermediate 
$\sigma$-resonance channel. The measured branching ratio of the two-body 
decay $D^+ \to \sigma \pi^+$ relative to the uncorrelated $3\pi$ decay of 
the $D$ meson is then deduced as 
$\Gamma(\sigma\pi^+)/\Gamma(\pi^+\pi^+\pi^-) = 
0.695 \pm 0.135 \pm 0.032$~\cite{PDG,E791}. 
The BES experiment~\cite{BES} concentrated on the nonleptonic decay 
$J/\Psi \to \sigma \omega \to \pi \pi \omega$. The extracted values of 
the $\sigma$ mass and width are: $M_\sigma = 390_{-36}^{+60}$ MeV and  
$\Gamma_\sigma = 282_{-50}^{+77}$ MeV. 

Preliminary analyses of these two experiments were performed 
in Refs.~\cite{Dib,Gatto,Huo}.
In Ref.~\cite{Dib} the relevant coupling constants of the respective
two-body decays $D\to\sigma\pi$ and $\sigma\to\pi\pi$ were extracted
from the data of the E791 experiment~\cite{E791}.
A direct calculation of the $D \to \sigma \pi$ amplitude was done in 
Ref.~\cite{Gatto} in a constituent quark-meson model. Both analyses neglect 
the intrinsic momentum dependence of the $D\to \sigma \pi$ transition form 
factor and, in the case of Ref.~\cite{Dib}, the final state interaction in 
the three-body decay $D\to 3\pi$. The two approaches~\cite{Dib,Gatto}   
arrive at a disagreement between the analysis of the nonleptonic 
two- and three-body decays of the $D$ meson. 
The extracted~\cite{Dib} or calculated~ \cite{Gatto} coupling constant 
$g_{D\sigma\pi}$ is approximately twice as large as the one deduced from 
experimental data on the two-body decay $D \to \sigma\pi$~\cite{PDG}. 
In Ref.~\cite{Huo} the effective coupling constant 
$g_{J/\psi\sigma\omega}$ was estimated using the perturbative QCD 
technique. Also, the role of the light $\sigma$ as a elementary 
particle~\cite{Deandrea} and as a correlated two-pion state~\cite{Gardner}
was examined in $B\to\rho\pi$ decay. 

In the present paper we consider the two-body nonleptonic decays involving 
the light $\sigma$-meson with $D \to \sigma\pi$ and $B \to \sigma\pi$. 
We work in the framework of a QCD motivated, relativistic quark model which 
implements a linear realization of chiral 
symmetry~\cite{Model}-\cite{EPJdirect}.
In this context the formalism also allows to describe the pion as a
composite particle. To solidify and extent our considerations,
we therefore also present a comprehensive analysis of 
the electromagnetic form factors of $\pi$ and $\sigma$
associated with the transitions $\pi \to \pi \gamma$,  
$\pi\to\gamma\gamma$ and $\sigma\to\gamma\gamma$. 

The specific scheme we work in can be viewed as an effective quantum field 
theory approach based on a Lagrangian of light and heavy hadrons 
(both mesons and baryons) interacting with their constituent 
quarks~\cite{Model}-\cite{EPJdirect}. 
The coupling strength of a specific hadron to its constituent quarks
is determined by the compositeness condition $Z_H=0$~\cite{Salam,Efimov},
where $Z_H$ is the wave function renormalization constant of the hadron.
The compositeness condition enables to relate theories with quark and hadron 
degrees of freedom to
effective Lagrangian approaches formulated in terms of hadron variables 
only (as, for example, Chiral Perturbation Theory~\cite{ChPT} and its 
covariant extension to the baryon sector~\cite{Becher_Leutwyler}). 

Our strategy is as follows. We start with an effective interaction
Lagrangian written down in terms of quark and hadron variables.
Then, by using Feynman rules, the $S$-matrix elements describing
hadron-hadron interactions are given by a set of quark diagrams.
The compositeness condition is sufficient to avoid double counting of 
quark and hadron degrees of freedom. The Lagrangian contains only 
a few model parameters: the masses of light and heavy quarks, and 
scale parameters which define the size of the distribution of 
the constituent quarks inside the hadron. This approach has been 
previously used to compute exclusive semileptonic, nonleptonic, 
strong and electromagnetic decays of light and heavy hadrons
\cite{Model}-\cite{EPJdirect} employing the same set of model parameters. 

In such a way we consider the $\sigma$-meson as a 
quark-antiquark bound state of the light $u$ and $d$ flavors.
We assume that the coupling strengths
of both the pion and the sigma meson to the constituent quarks are
identical in accordance with the linear realization of chiral symmetry.
Based on this scheme the $\sigma$ meson mass is determined as
$M_\sigma=385.4$ MeV, which is in good agreement with the BES-result.
Next we discuss in detail the gauging of the nonlocal meson-quark
interaction and, for consistency, calculate the electromagnetic form factor 
of the pion. We then proceed with the $\pi^0\to\gamma\gamma$
and $\sigma\to\gamma\gamma$ processes. Here we obtain explicit
expressions for the relevant form factors and evaluate
the leading and next-to-leading order for large Euclidean  
photon virtualities. As a result we also obtain the two-photon
decay width of the $\sigma$-meson. We compute the strong decay
width for the process $\sigma\to\pi\pi$ and finally turn to the
nonleptonic decays $D\to\sigma\pi$ and $B\to\sigma\pi$.
All our results are compared to available experimental data
and to other theoretical studies.

The layout of the paper is as follows. In Sec.~II we begin by
introducing the relativistic quark model which implements
a linear realization of chiral symmetry.
Sec.~III is devoted to the derivation of the electromagnetic properties
of the $\pi$ and $\sigma$ mesons. 
In Sec.~IV we examine the strong and nonleptonic decay characteristics of
the transitions $\sigma\to\pi\pi$ and $D(B)\to\sigma\pi$. 
Then in Sec.~V we turn to a numerical analysis of the processes
considered and discuss the quality of results in comparison
with experiment and previous model calculations.  
Finally, we summarize our results in Sec.~VI. 

\section{The model}\label{sec:model}

We will consistently employ the relativistic constituent quark 
model~\cite{Model}-\cite{EPJdirect} to compute a variety of
observables related to the $\pi$ and $\sigma$ mesons.
In the following we will present details of the model which is
essentially based on an effective interaction Lagrangian describing
the coupling between hadrons and their constituent quarks.

The coupling of a meson $H(q_1\bar q_2)$ to its constituent 
quarks $q_1$ and $\bar q_2 $ is set up by the Lagrangian 
\begin{equation}\label{Lagr_str}
{\cal L}_{\rm int}^{\rm str}(x) = g_H H(x)\int\!\! dx_1 \!\!\int\!\! 
dx_2 F_H (x,x_1,x_2)\bar q_2(x_2)\Gamma_H\lambda_H q_1(x_1) \, + {\rm h.c.}  
\end{equation}
Here, $\lambda_H$ and $\Gamma_H$ are  Gell-Mann and Dirac matrices 
which inclose the flavor and spin quantum numbers of the meson field 
$H(x)$. The function $F_H$ is related to the scalar part of the 
Bethe-Salpeter amplitude and characterizes the finite size of the 
meson. To satisfy translational invariance the function $F_H$
has to fulfil the identity $F_H(x+a,x_1+a,x_2+a)=F_H(x,x_1,x_2)$ for
any 4-vector a. In the following we use 
a particular form for the vertex function 
\begin{equation}\label{vertex}
F_H(x,x_1,x_2)=\delta(x - w_{21}x_1 - w_{12}x_2) \Phi_H((x_1-x_2)^2) 
\end{equation}
where $\Phi_H$ is the correlation function of two constituent quarks 
with masses $m_1$, $m_2$ and $w_{ij}=m_j/(m_i+m_j)$.  

The coupling constant $g_H$ in Eq.~(\ref{Lagr_str}) is determined by the 
so-called {\it compositeness condition} originally proposed in~\cite{Salam}, 
and extensively used in~\cite{Model}-\cite{EPJdirect},\cite{Efimov}.  
The compositeness condition requires that the renormalization constant of 
the elementary meson field $H(x)$ is set to zero 
\begin{equation}\label{z=0}
Z_H \, = \, 1 - \, \frac{3g^2_H}{4\pi^2} \, 
\tilde\Pi^\prime_H(M^2_H) \, = \, 0
\end{equation}
where $\tilde\Pi^\prime_H$ is the derivative of the meson mass operator. 
To clarify the physical meaning of this condition, we first want to remind 
that the renormalization constant $Z_H^{1/2}$ is also interpreted as  
the matrix element between the physical and the corresponding bare state.  
For $Z_H=0$ it then follows that the physical state does not contain  
the bare one and is described as a bound state. 
The interaction Lagrangian of Eq.~(\ref{Lagr_str}) and 
the corresponding free parts describe on the same level 
both the constituents (quarks) and the physical particles (hadrons)
which are supposed to be the bound states of the constituents.
As a result of the interaction, the physical particle is dressed, 
i.e. its mass and wave function are to be renormalized. 
The condition $Z_H=0$ also effectively excludes
the constituent degrees of freedom from the physical space
and thereby guarantees that double counting
for the physical observable under consideration is avoided.
Now the constituents exist in virtual states only. 
One of the corollaries of the compositeness condition is the absence 
of a direct interaction of the dressed charged particle with the 
electromagnetic field. Taking into account both the tree-level
diagram and the diagrams with the self-energy insertions into the 
external legs (that is the tree-level diagram times $Z_H -1$) yields 
a common factor $Z_H$  which is equal to zero. We refer the interested 
reader to our previous papers \cite{Model}-\cite{EPJdirect},\cite{Efimov}
where these points are discussed in great details. 

The meson mass operator appearing in Eq.~(\ref{z=0}) is 
described by the Feynman diagram in Fig. 1. In the case of pseudoscalar 
$(\Gamma_H = i\gamma^5)$ and scalar $(\Gamma_H = I)$ mesons, relevant 
for the present paper, we obtain the expression 
\begin{eqnarray}\label{Mass-operator}
\tilde\Pi^\prime_H(p^2) \, = \, - \, \frac{p^\alpha}{2p^2}\,
\frac{d}{dp^\alpha}\,
\int\!\! \frac{d^4k}{4\pi^2i} \tilde\Phi^2_H(-k^2) 
{\rm tr} \biggl[\Gamma_H S_1(\not\! k+w_{21} \not\!p) 
\Gamma_H S_2(\not\! k-w_{12} \not\!p) \biggr] \, ,  
\end{eqnarray}
where $\tilde\Phi_H(-k^2)$ is the Fourier-transform of the 
correlation function  $\Phi_H((x_1-x_2)^2)$ and 
$S_i(\not\! k)$ is the quark propagator. 
We use free fermion propagators for the valence quarks 
\begin{equation}
S_i(\not\! k)=\frac{1}{m_i-\not\! k} 
\end{equation}
with an effective constituent quark mass $m_i$. As discussed 
in~\cite{Model}-\cite{EPJdirect} we assume for the meson mass $M_H$ that 
\begin{equation}
\label{conf}
M_H < m_{1} + m_{2}
\end{equation}
in order to avoid the appearance of imaginary parts in the 
physical amplitudes. The calculational technique for determining the 
explicit expression of $\tilde\Pi^\prime_H(p^2)$~(\ref{Mass-operator}) is 
outlined in Appendix~A. 

The interaction with the electromagnetic field is introduced
by two ways. The free Lagrangians of quarks and hadrons
are gauging in a standard manner by using minimal
substitution:

\begin{equation}
\label{photon}
\partial^\mu H^\pm \to (\partial^\mu \mp ie A^\mu) H^\pm ,
\hspace{1cm} 
\partial^\mu q \to (\partial^\mu - ie A^\mu) q, 
\hspace{1cm} 
\partial^\mu \bar q \to (\partial^\mu +ie A^\mu)\bar q,
\end{equation}
where $e$ is a proton charge. It gives us the first
piece of the electromagnetic interaction Lagrangian
written as

\begin{eqnarray}
\label{L_em_1}
{\cal L}^{\rm em (1)}_{\rm int}(x) &=& 
e \bar q(x) \not\! A Q \, q(x) 
\\
&+&ie A_\mu(x)\left(H^-(x)\partial^{\,\mu} H^+(x)
                 -H^+(x)\partial^{\,\mu} H^-(x)\right)
+ \, e^2  A_\mu^2(x) H^-(x) H^+(x).   
\nonumber
\end{eqnarray}

The gauging nonlocal Lagrangian in Eq.~(\ref{Lagr_str})
proceeds in a way suggested in \cite{Terning}.
To guarantee local invariance of the strong interaction Lagrangian, in  
${\cal L}_{\rm int}^{\rm str}$ one multiplies 
each quark field $q(x_i)$ with the gauge field exponentional
that gives

\begin{eqnarray}
\label{gauging}
{\cal L}_{\rm int}^{\rm str + em(2)}(x) &=&
g_H H(x)\int\!\! dx_1 \!\!\int\!\!dx_2 F_H (x,x_1,x_2)
\bar q_2(x_2)\, e^{ie_{q_2} I(x_2,x,P)}\, \\
&\times&\Gamma_H\lambda_H\,  
e^{-ie_{q_1} I(x_1,x,P)}\,  q_1(x_1),  \nonumber 
\end{eqnarray}
where

\begin{equation}
\label{path}
I(x_i,x,P) = \int\limits_x^{x_i} dz_\mu A^\mu(z). 
\end{equation}

It is readily seen that the full Lagrangian is invariant 
under transformations

\begin{eqnarray*}
q_i(x) &\to& e^{ie_{q_i} f(x)} q_i(x),    \hspace{1cm}
\bar q_i(x) \to \bar q_i(x) e^{-ie_{q_i} f(x)}, \hspace{1cm}
H(x)\to e^{ie_{H} f(x)}H(x), 
\\
A^\mu(x)&\to& A^\mu(x)+\partial^\mu f(x),
\end{eqnarray*}
where $e_H=e_{q_2}-e_{q_1}$.

Then the second term of the electromagnetic interaction Lagrangian
${\cal L}^{em}_{\rm int; 2}$ arises, when expanding the gauging exponential
up to a certain power of $A_\mu$, relevant for the order of perturbation 
theory and for the process we consider. 
It seems that the results will be dependent on the path $P$
which connects the end-points in the path integral in Eq~(\ref{path}).
However, we need to know only derivatives of such
integrals under calculations within the perturbative series.
Therefore, we use a formalism~\cite{Model,Mandelstam,Terning} 
which is based on the path-independent definition of the derivative of 
$I(x,y,P)$: 
\begin{eqnarray}\label{path1}
\lim\limits_{dx^\mu \to 0} dx^\mu 
\frac{\partial}{\partial x^\mu} I(x,y,P) \, = \, 
\lim\limits_{dx^\mu \to 0} [ I(x + dx,y,P^\prime) - I(x,y,P) ]
\end{eqnarray}
where path $P^\prime$ is obtained from $P$ when shifting the end-point $x$
by $dx$.
Use of the definition (\ref{path1}) 
leads to the key rule
\begin{eqnarray}\label{path2}
\frac{\partial}{\partial x^\mu} I(x,y,P) = A_\mu(x)
\end{eqnarray}
which in turn states that the derivative of the path integral $I(x,y,P)$ does 
not depend on the path P originally used in the definition. The non-minimal 
substitution (\ref{gauging}) is therefore completely equivalent to the 
minimal prescription as evident from the identities (\ref{path1}) or 
(\ref{path2}). The method of deriving Feynman rules for a non-local coupling 
of hadrons to photons and quarks was already developed in 
Refs.~\cite{Model,Terning} and will be discussed in the next section, where 
we apply the formalism to the physical processes considered here. 

For example, the piece of the Lagrangian in Eq.~(\ref{gauging})
in the first order over charge reads as

\begin{eqnarray}\label{first}
{\cal L}_{\rm int}^{\rm em(2)}(x) &=&   
g_H H(x)\int\!\! dx_1 \!\!\int\!\!dx_2\int\!\!dy\, 
E^\mu_H(x,x_1,x_2,y)\,  A_\mu(y)\, 
\bar q_2(x_2)\Gamma_H\lambda_H q_1(x_1)\,,
\\
&&\nonumber\\ 
E^\mu_H(x,x_1,x_2,y) &=&
\int\frac{dp_1}{(2\pi)^4}\int\frac{dp_2}{(2\pi)^4}
\int\frac{dq}{(2\pi)^4}
e^{ip_1(x_1-x)-ip_2(x_2-x)+iq(y-x)}\tilde E^\mu_1(p_1,p_2,q)\,,
\nonumber\\
&&\nonumber\\ 
\tilde E^\mu_1(p_1,p_2,q) &=& -e_{q_1}w_{12}(w_{12} q^\mu+2 p_0^\mu)
\int\limits_0^1 dt \tilde\Phi'_H\left(-t(w_{12}q+p_0)^2-(1-t)p_0^2\right)
\nonumber\\
&+&
e_{q_2}w_{21}(w_{21} q^\mu-2  p_0^\mu)
\int\limits_0^1 dt \tilde\Phi'_H\left(-t(w_{21}q-p_0)^2-(1-t)p_0^2\right)\,,
\nonumber\\
&&\nonumber\\ 
p_0 &=& w_{12}p_1+w_{21}p_2.
\end{eqnarray}

Transition matrix elements involving composite hadrons are specified in the
model by the appropriate quark diagram.
For example, the transition form factor which characterizes the hadronic 
transition $H_{13}(p_1)  \to  H_{23}(p_2) + H_{12}(p_3)$ 
is determined from the Feynman integral corresponding to 
the diagram of Fig.~2: 
\begin{eqnarray}\label{Lambda_Trans} 
\Lambda^{12; 13; 23}(p_1,p_2) &=&
\frac{3}{4\pi^2} g_{H_{13}} g_{H_{23}} g_{H_{12}} 
I^{12; 13; 23}(p_1,p_2) \nonumber\\ 
I^{12; 13; 23}(p_1,p_2)  &=& - \, \int\!\! \frac{d^4k}{4\pi^2i} \, 
\tilde\Phi_{H_{13}}( - (k+w_{13}\,p_1)^2) \, 
\tilde\Phi_{H_{23}}( - (k+w_{23}\,p_2)^2)\\
&\times&\tilde\Phi_{H_{12}}( - (k+w_{12}\,p_1+w_{21}\,p_2)^2) 
{\rm tr} [ S_2(\not\! k+\not\! p_2) \Gamma_{H_{12}}  
S_1(\not\! k+\not\! p_1)\Gamma_{H_{13}} 
S_3(\not\! k)\Gamma_{H_{23}} ]\nonumber
\end{eqnarray}
where technical details concerning the derivation of integral 
(\ref{Lambda_Trans}) are indicated in Appendix~A.

Finally we have to specify the vertex function 
$\tilde\Phi_H$ (Eq. (\ref{Mass-operator})), which characterizes 
the finite size of the hadrons. 
Any choice for $\tilde\Phi_H$ is appropriate  
as long as it falls off sufficiently fast in the ultraviolet region of 
Euclidean space to render the Feynman diagrams ultraviolet finite. 
We employ a Gaussian for the vertex function 
$\tilde\Phi_H(k^2_E) \doteq \exp(- k^2_E/\Lambda^2_H)$, where $k_E$ is 
an Euclidean momentum. The size parameters $\Lambda^2_H$ are determined 
by fitting to experimental data, when available, or to lattice results 
for the leptonic decay constants $f_P$ where 
$P=\pi, D, B$.  
The leptonic decay constant $f_P \doteq {\cal F}_P(M_P^2)$ is determined
from \cite{EPJdirect}
\begin{equation}
\label{leptonic}\label{fP} 
{\cal F}_P(p^2) \,p^\mu \, = \, \frac{3g_P}{4\pi^2} 
\,\int\!\! \frac{d^4k}{4\pi^2i} 
\tilde\Phi_P(-k^2) {\rm tr} \biggl[O^\mu S_1(\not\! k+w_{21} \not\!p) 
\gamma^5 S_2(\not\! k-w_{12} \not\!p) \biggr] \, .
\end{equation}  
To reduce the set of values for $\Lambda_H$, we use a unified size 
parameter for hadrons with the same flavor content. 
The best fit to the decay constants $f_P$ is obtained (see Table I) 
when the values of the constituent quark masses and the parameters 
$\Lambda_H$ are choosen as follows \cite{EPJdirect} 
\begin{equation}
\begin{array}{ccccc}
m_{u(d)} & m_s & m_c & m_b & \\  \hline
$\ \ 0.235\ \ $ & $\ \ 0.333\ \ $ & $\ \ 1.67\ \ $ & $\ \ 5.06\ \ $
&
$\ \ {\rm GeV} $\\
\end{array}\label{fitmas}
\end{equation}
and
\begin{equation}
\begin{array}{ccccc}
\Lambda_{\pi(\sigma)} & \Lambda_K & \Lambda_D & \Lambda_B & \\  \hline
$\ \ 1 \ \ $ & $\ \ 1.6\ \ $ & $\ \ 2\ \ $ & $\ \ 2.25\ \ $ &
$\ \ {\rm GeV} $\\
\end{array}\label{fitlambda}
\end{equation}

In this paper we are aiming to explore the properties
of pions and light $\sigma$-meson. We write down the
relevant interaction Lagrangian   
in accordance with the two-flavor linear 
$\sigma$-model~\cite{Schwinger,Gell-Mann_Levy}  
\begin{eqnarray}\label{sigma_model}
{\cal L}_{\rm int}^{\rm str; \pi + \sigma}(x) \, = \, \frac{g}{\sqrt{2}} \, 
\int\!\! dy \,  \Phi(y^2) \, \bar q(x + y/2) \, [ \, \sigma(x) \, + 
\, i \gamma^5 \, \vec{\pi}(x) \, \vec{\tau} \, ] \, q(x - y/2) \,. 
\end{eqnarray}
Here the constraints on the bare couplings  
$g \doteq g_{\pi} \equiv g_{\sigma}$ and
the vertex functions $\Phi\doteq\Phi_{\pi}\equiv\Phi_{\sigma}$ 
are imposed by chiral symmetry. 
We also require that the dressed couplings determined
from  the compositeness condition of Eq.(\ref{z=0})
should be equal each other.
This requirement allows us to determine the $\sigma$ meson mass by 
the equality 
\begin{eqnarray}\label{Matching_sigma}
\tilde\Pi^\prime_\sigma(M^2_\sigma) \, = \,
\tilde\Pi^\prime_\pi(M^2_\pi) \, ,
\end{eqnarray}
where the physical pion mass $M_{\pi}$ is used as an input.

With the hadron parameters fixed, from Eq.~(\ref{Matching_sigma}), 
we deduce a $\sigma$ mass of
\begin{equation}
M_\sigma = 385.4 \, {\rm MeV}~.
\end{equation}
The predicted value we obtain is close to the BES result~\cite{BES} of 
$M_\sigma = 390_{-36}^{+60}$ MeV, which sets the lower scale for the range 
of mass values compiled in~\cite{PDG}.

\section{Electromagnetic properties of $\pi$ and $\sigma$}

In this section we proceed with the formalism on the
electromagnetic properties of $\pi$ and $\sigma$ mesons.
For consistency we also include the $\pi$ meson in the discussion,
where, in addition, the model predictions will serve to
solidify the validity of the previously outlined approach.
In the analysis we consider the following related processes:
the electromagnetic transition of
charged pions, that is $\pi^\pm \to \pi^\pm \gamma$, and
the form factors characterizing 
the transitions $\pi^0\to\gamma\gamma$ and $\sigma\to\gamma\gamma$ 
for different kinematical regimes of the photons (real and virtual)
All amplitudes considered are obtained in a manifestly 
gauge-invariant form. 

When gauging (\ref{gauging}) the nonlocal strong interaction Lagrangian
${\cal L}_{\rm int}^{\rm str}$, additional "contact" vertices are generated,
which couple hadrons, quarks and photons, as already discussed in Sec.~II.
In particular, for the process $\pi^\pm \to \pi^\pm \gamma$ we need a vertex 
describing the coupling of a charged pion, a single photon and two quarks 
(Fig.3a). The coupling of $\sigma$ to a photon and two quarks (Fig.3a) and 
to two photons and two quarks (Fig.3b) will contribute to the transition 
$\sigma\to\gamma\gamma$.

The full set of Feynman diagrams for the electromagnetic transition amplitudes
considered are summarized in the following: The transition 
$\pi^\pm \to \pi^\pm \gamma$ is described by a triangle diagram (Fig.4a) 
and the two additional contact diagrams of Figs.4b and 4c. 
As mentioned in Sec.II, the diagram describing the direct coupling of pions 
to a photon is compensated by the counterterm (see Eq.~(\ref{L_em_1}). 
The process $\pi^0\to\gamma\gamma$ obtains a contribution by the single diagram of 
Fig.5a. The transition amplitude $\sigma\to\gamma\gamma$ is generated by
the four diagrams of Fig.5: a triangle diagram (Fig.5a) and three contact
diagrams (Figs.5b-d).  
In the following we denote diagrams containing a contact vertex with a
single photon line (Figs.4b, 4c, 5b and 5c) as "bubble"-diagrams
and the amplitude of Fig.5d as "tadpole"-diagram. The dominant 
contribution to the electromagnetic form factors arises from the
leading triangle diagrams (Fig.4a and 5a). We will demonstrate
that diagrams containing contact vertices
give negligible contribution, but in general they should be kept to guarantee 
electromagnetic gauge invariance. Feynman rules for the evaluation of
the  nonlocal vertices of Figs.3a and 3b are derived in Appendix~B. Next we 
will give the definition of the amplitudes for the processes 
$\pi^\pm \to \pi^\pm \gamma$, $\pi^0\to\gamma\gamma$ and 
$\sigma\to\gamma\gamma$ and discuss the properties of the relevant 
form factors. 

\subsection{The electromagnetic form factor $F_{\pi}(Q^2)$ of the pion} 

It is convenient to write down the vertex function for the transition 
$\pi^\pm(p)\to\pi^\pm(p^\prime) \, + \, \gamma(q)$ in the form 
\begin{eqnarray}\label{vertex_pion}
\Lambda^\mu(p,p^\prime) = \frac{q^\mu}{q^2} [\tilde\Sigma_\pi(p^2) - 
\tilde\Sigma_\pi(p^{\prime \, 2})] \, + \, \Lambda^\mu_\perp(p,p^\prime) 
\end{eqnarray} 
from which the Ward-Takahashi identity follows immediately 
\begin{eqnarray}
q_\mu \Lambda^\mu(p,p^\prime) = \tilde\Sigma_\pi(p^2) - 
\tilde\Sigma_\pi(p^{\prime \, 2}) \, . 
\end{eqnarray}
For setting up the electromagnetic vertex function of Eq. (\ref{vertex_pion})
we use the pion mass operator $\tilde\Sigma_\pi(p^2)$ with 
\begin{eqnarray}
\tilde\Sigma_\pi(p^2) &\doteq&\frac{3g^2}{4\pi^2}\,\tilde\Pi_\pi(p^2)\,,\\ 
\tilde\Pi_\pi(p^2) &=& \int \frac{d^4k}{4\pi^2 i} \, 
\tilde\Phi(-k^2) \, D(p) \,,\,\,\,\,\,\,\,\,\,  
D(p) \, = \, {\rm tr}\biggl[\gamma^5 S(k + \frac{p}{2}) 
\gamma^5 S(k + \frac{p}{2}) \biggr] \,  \nonumber
\end{eqnarray} 
and $\Lambda^\mu_\perp(p,p^\prime)$ is the part
which is orthogonal to the photon momentum: 
$q_\mu \, \Lambda^\mu_\perp(p,p^\prime) = 0$. The explicit expression for 
$\Lambda^\mu_\perp(p,p^\prime)$ results from the sum of the gauge-invariant 
parts of the triangle $(\Delta)$ (Fig.4a) and of the
bubble (bub) diagrams (Figs.4b and 4c): 
\begin{eqnarray}
\Lambda^\mu_\perp(p,p^\prime)\, = \,\Lambda^\mu_{\triangle_\perp}(p,p^\prime) 
\, + \, \Lambda^\mu_{\rm bub_\perp}(p,p^\prime); \hspace{1cm} 
\Lambda^\mu_{\triangle_\perp ({{\rm bub}_\perp})}(p,p^\prime) \, = \, 
\frac{3g^2}{4\pi^2}\, I^\mu_{\triangle_\perp(\rm bub_\perp)}(p,p^\prime)\, . 
\end{eqnarray}
The separate contributions $I^\mu_{\triangle_\perp}(p,p^\prime)$ 
and $I^\mu_{\rm bub_\perp}(p,p^\prime)$ are given by
\begin{eqnarray}\label{I_part}
I^\mu_{\triangle_\perp}(p,p^\prime) &=&  \int \frac{d^4k}{4\pi^2 i} \, 
\tilde\Phi\biggl(-\biggl[k+\frac{p}{2}\biggr]^2\biggr) \, 
\tilde\Phi\biggl(-\biggl[k+\frac{p^\prime}{2}\biggr]^2\biggr) \, 
{\rm tr}[\gamma^5 S(k + p^\prime) \gamma_{\perp; \, q}^\mu 
S(k + p) \gamma^5 S(k)] \,,\nonumber\\ 
I^\mu_{\rm bub_\perp}(p,p^\prime) &=& \frac{\eta^\mu}{\eta^2} \, 
\int \frac{d^4k}{4\pi^2 i} \, \tilde\Phi(-k^2) \, \int\limits_0^1 dt \, 
\tilde\Phi^{\, \prime}\biggl(- k^2 - t\biggl[kq + \frac{q^2}{4}\biggr]\biggr) 
\,\, k\eta \,\, \biggl\{ D(p) -  D(p^\prime) \biggr\} \,. 
\end{eqnarray}
where $\tilde\Phi^{\, \prime}(z) = \partial\tilde\Phi(z)/\partial z$. Also,
$\gamma^\mu_{\perp; \, q} = \gamma^\mu  -  q^\mu\not\! q/q^2$ 
and $\eta^\mu  =  P^\mu  -  (Pq/q^2)  q^\mu$ are  
orthogonal to the momentum transfer $q$ with $P \, = \, p \, + p^\prime$.

In the limit $q=0$ we obtain
\begin{eqnarray}\label{Ward_id}
\Lambda^\mu(p,p) \, = \, 
\frac{\partial\tilde\Sigma_\pi(p^2)}{\partial p^\mu} \, = \, 
2 \, p^\mu \, \frac{\partial\tilde\Sigma_\pi(p^2)}{\partial p^2} \, , 
\end{eqnarray}
while, by definition, we also have
\begin{eqnarray}\label{normaliz}
\Lambda^\mu(p,p) = 2 p^\mu F_{\pi} (0) 
\end{eqnarray}  
where $F_{\pi}(0)$ is the pion charge form factor at the 
origin. From the comparison of Eqs. (\ref{Ward_id}) and 
(\ref{normaliz}) it follows that the compositeness condition (\ref{z=0}) 
is equivalent to the normalization of the pion form factor at the 
origin with $F_{\pi}(0)=1$. 

The full electromagnetic form factor $F_{\pi}(Q^2)$ of the pion is defined by 
\begin{eqnarray}
\Lambda^\mu(p,p^\prime)\biggr|_{p^2 = p^{\prime \, 2} = M_{\pi}^2} = P^\mu \, 
F_{\pi}(Q^2) 
\end{eqnarray}
where $Q^2 = - q^2 = - (p - p^\prime)^2$ is the Euclidean momentum transfer 
squared. The electromagnetic radius of the pion, related to the slope of 
$F_{\pi}(Q^2)$ at the origin,  is then given by 
\begin{eqnarray}
<r^2_{\pi}> = - 6 \frac{dF_{\pi}(Q^2)}{dQ^2}
\Bigg|_{\displaystyle{Q^2 = 0}} \, . 
\end{eqnarray}

\subsection{Transitions $\pi^0\to\gamma\gamma$ and 
$\sigma\to\gamma\gamma$} 

The matrix elements for the transitions $\pi^0\to\gamma\gamma$ 
and $\sigma\to\gamma\gamma$ can be written in general in the 
manifestly gauge-invariant form  
\begin{eqnarray}
M_{\pi^0 \to \gamma \gamma}^{\mu\nu}(q_1,q_2) &=& e^2 \, 
\, a^{\mu\nu} \, F_{\pi\gamma\gamma}(p^2,q_1^2,q_2^2) \, , \label{axial_an}\\
M_{\sigma \to \gamma \gamma}^{\mu\nu}(q_1,q_2) &=& e^2 \, \biggl\{ \, 
b^{\mu\nu} \, F_{\sigma\gamma\gamma}(p^2,q_1^2,q_2^2) 
\, + \, c^{\mu\nu} \, G_{\sigma\gamma\gamma}(p^2,q_1^2,q_2^2)\,\biggr\}\, .  
\label{scalar_an} 
\end{eqnarray} 
The tensors $a^{\mu\nu}$, $b^{\mu\nu}$ and $c^{\mu\nu}$ refer to the 
Lorentz structures: 
\begin{eqnarray}
a^{\mu\nu}&=&\epsilon^{\mu\nu\alpha\beta}\,(q_1)_\alpha\, (q_2)_\beta \,\,\,,
\nonumber\\
b^{\mu\nu} &=& g^{\mu\nu} \, (q_1q_2) \, - \, q_1^\nu q_2^\mu \,\,\,,\\
c^{\mu\nu} &=& g^{\mu\nu} \, q_1^2 q_2^2 \, + \, q_1^\mu q_2^\nu \, (q_1q_2) 
\, - \, q_1^\mu q_1^\nu \, q_2^2 \, - \, q_2^\mu q_2^\nu \, q_1^2 \,\,\,,
\nonumber
\end{eqnarray}
while $q_1$ and $q_2$ are the photon four-momenta and $p = q_1 + q_2$ is the 
meson momentum. Here, $F_{\pi\gamma\gamma}(p^2,q_1^2,q_2^2)$ is the 
$\pi^0\to\gamma\gamma$ form factor. The transition $\sigma\to\gamma\gamma$ is 
characterized by two form factors: $F_{\sigma\gamma\gamma}(p^2,q_1^2,q_2^2)$ and 
$G_{\sigma\gamma\gamma}(p^2,q_1^2,q_2^2)$. Usually only 
one of the form factors, that is $F_{\sigma\gamma\gamma}$, is discussed in 
the literature. However, when both photons are off-shell, constraints set by 
gauge invariance result in the two terms of the matrix element 
$M_{\sigma \to \gamma \gamma}^{\mu\nu}$, 
proportional to the Lorentz structures $b^{\mu\nu}$ and $c^{\mu\nu}$. 
If at least one of the photons is real, then only 
the first form factor $F_{\sigma\gamma\gamma}$ gives a non-vanishing
contribution to the invariant 
matrix element. The decay width of the transition $H \to \gamma\gamma$ 
with $H = \pi$ or $\sigma$ is given by 
\begin{eqnarray}\label{dec_H} 
\Gamma_{H\gamma\gamma} \, = \, \frac{\pi}{4} \, \alpha^2 \, M_H^3 \, 
g_{H\gamma\gamma}^2  
\end{eqnarray}
where $g_{H\gamma\gamma} \, = \, F_{H\gamma\gamma}(M_H^2,0,0)$ is the 
$\pi\gamma\gamma$ coupling constant and $\alpha = e^2/(4\pi)$.  

For the evaluation of the $H\to\gamma\gamma$  
form factors we pursue the following strategy. First we consider the 
calculation of the form factors $F_{\pi\gamma\gamma}(p^2,q_1^2,q_2^2)$, 
$F_{\sigma\gamma\gamma}(p^2,q_1^2,q_2^2)$ and 
$G_{\sigma\gamma\gamma}(p^2,q_1^2,q_2^2)$ 
in so-called local limit. In this case
the dimensional parameter $\Lambda_\pi$, appearing in the 
$\pi (\sigma)$ meson vertex function $\tilde\Phi(-k^2)$,
is taken to infinity or, which is equivalent, 
$\tilde\Phi(-k^2) \to 1$. We obtain analytical expressions for the 
$H\to\gamma\gamma$ form factors and show that for the case of 
$\pi^0\to\gamma\gamma^\ast$
the form factor has the incorrect behavior for large photon virtualities
in the Euclidean region. Then we repeat the derivation using the nonlocal 
approach and show that the correct asymptotics for the $\pi\gamma\gamma^\ast$
form factor can be reproduced when using the mesonic quark-antiquark wave 
function. The final numerical analysis for the complete form factors will be 
done in a separate section. 

\subsubsection{Local $\pi^0\gamma\gamma$ and $\sigma\gamma\gamma$ diagrams} 

The Feynman integral corresponding to the local (L) triangle diagram is written as

\begin{eqnarray}\label{triangle} 
I^{\mu\nu; \, M}_{\triangle, \, \rm L}(q_1,q_2) 
&=& \int\!\frac{d^4 k}{4\pi^2 i}\,
{\rm tr}\bigg\{\gamma^\mu S(\not\! k+\not\! q_1) 
\Gamma_M (\not\! k-\not\! q_2) \gamma^\nu S(\not\! k)\bigg\}
\end{eqnarray}
where $\Gamma_\pi =i\gamma^5$ for $\pi$ and $\Gamma_\sigma = I$ for $\sigma$.
For simplicity we drop a common factor of $- 3/4\pi^2$. The complete 
integral (\ref{triangle}) is free of the logarithmic ultraviolet (UV) 
divergence. Separate contributions to Eq. (\ref{triangle}) with the UV 
divergences are treated using dimensional regularization~\cite{Dim_Reg}. In 
this case we use the definition for $\gamma^5$ in D dimensions (see details 
in~\cite{Yndurain}) to guarantee the fulfilment of two important properties: 
$(\gamma^5)^2 = I$ and ${\rm tr}(\gamma^5 \gamma^\mu \gamma^\nu) = 0$.  
For the pseudoscalar triangle the UV divergency is completely gone due to the 
contraction of the divergent integral with the trace 
${\rm tr}(\gamma^5 \gamma^\mu \gamma^\nu)$, which 
by definition is zero in any dimension. In the case of the $\sigma\gamma\gamma$ diagram 
the use of the gauge-invariant regularization 
(e.g., dimensional regularization) guarantees the cancellation of the UV 
parts and generates the correct finite part. 
The finite part of the $\sigma\gamma\gamma$ diagram is the 
same for both types of gauge-invariant regularizations. 
The master formulas used in dimensional regularization are
\begin{eqnarray*}
{\rm tr}(\gamma^5 \gamma^\mu \gamma^\nu) \,\, \int &d^D k& \,\, 
\frac{k^2}{(m^2 - k^2 - D_0)^3} = 0 \,\,\,\,\,\,\,\, \hspace*{2.2cm} 
({\rm for} \,\,\, \pi \,\,\, {\rm meson})\\ 
\int &d^D k& \,\, \frac{4 k^\mu k^\nu + g^{\mu\nu} (m^2 - k^2 - D_0)}
{(m^2 - k^2 - D_0)^3} = 0 \,\,\,\,\,\,\,\,\,\,\, ({\rm for} \,\,\, \sigma \,\,\, {\rm meson})\\
\end{eqnarray*}
where $D_0=\alpha_1\alpha_2 p^2+\alpha_1\alpha_3 q_1^2 + 
\alpha_2\alpha_3 q_2^2$ 
with $p=q_1+q_2$ arises when using the $\alpha$-parametrization.  

The integral Eq.(\ref{triangle}) is gauge invariant
$$
I^{\mu\nu; \, \Gamma}_{\triangle,\,\rm L}(q_1,q_2)\cdot (q_1)_\mu =
I^{\mu\nu; \, \Gamma}_{\triangle,\,\rm L}(q_1,q_2)\cdot (q_2)_\nu = 0
$$ 
as can be easily deduced from the Ward identity 
$$
S(\not\! k)\not\! q_1 S(\not\! k+\not\! q_1)=
S(\not\! k+\not\! q_1)-S(\not\! k)
$$
and simple algebra.

The integral of Eq.(\ref{triangle}) is particularly simple 
for the case of the pion with
\begin{eqnarray}\label{pion} 
I^{\mu\nu; \, \pi}_{\triangle,\,\rm L}(q_1,q_2) 
\, = \, m \, a^{\mu\nu} \, F^{\pi}_{\rm L}(p^2,q_1^2,q_2^2)
\end{eqnarray}
where the form factor $F^{\pi}_{\rm L}$ is given by the two-dimensional 
integral $I_{\rm L}$:  
\begin{eqnarray}\label{I_L} 
F^{\pi}_{\rm L}(p^2,q_1^2,q_2^2) \equiv I_{\rm L}(p^2,q_1^2,q_2^2) = 
\int\! d^3\alpha \, 
\delta\left(1-\sum\limits_{i=1}^3 \alpha_i\right)\,
\frac{1}{m^2-D_0}\,. 
\end{eqnarray}
The double integral $I_{\rm L}$ can be further reduced to a single one 
using the 't Hooft-Veltman technique \cite{Hooft} and can be finally 
expressed by a combination of Spense functions. Here, however, 
we will stay with the integral representation. We have 
\begin{eqnarray}\label{triangle_loc} 
&&
I_{\rm L}(p^2,q_1^2,q_2^2) =
\int\limits_0^1\! dx\,
\left\{
\frac{\ln[1-x(1-x)\,p^2/m^2]}
{\lambda^{1/2}\cdot x+q_1^2-\alpha_0\,p^2}
\right.
\\
&&\nonumber\\
&&
\left. - \,\frac{\ln[1-x(1-x)\,q_2^2/m^2]}
{\lambda^{1/2}\cdot x-q_1^2-q_2^2\,\alpha_0/(1-\alpha_0)}
+\,\frac{\ln[1-x(1-x)\,q_1^2/m^2]}
{\lambda^{1/2}\cdot x+q_1^2\,(1-\alpha_0)/\alpha_0+q_2^2}
\right\}\, ,
\nonumber
\end{eqnarray}
where 
\begin{eqnarray}\label{lambda_tri}
\lambda \doteq \lambda(p^2,q_1^2,q_2^2) = 
p^4 + q_1^4 +q_2^4 - 2 p^2 q_1^2 - 2 p^2 q_2^2 - 2 q_1^2 q_2^2 
\end{eqnarray}
is a kinematical triangle function and 
$\alpha_0=(p^2+q_1^2-q_2^2+\lambda^{1/2})/2p^2$.
All three thresholds can be readily seen from this
representation. In particular, when both photons are on their mass-shell
($q_1^2=q_2^2=0$) we get

\begin{eqnarray}\label{triangle_shell} 
I_{\rm L}(p^2,0,0) &=&
-\,\frac{1}{2 p^2}\int\limits_0^1\!  dx\,\,
\frac{\ln[1-x(1-x)\,p^2/m^2]}{x(1-x)}
\\
&&\nonumber\\
&=&
\frac{1}{4}\int\limits_0^1\! dv\,
\ln\left(\frac{1+\sqrt{1-v}}{1-\sqrt{1-v}}\right)
\cdot \frac{1}{m^2-v\cdot p^2/4-i\epsilon}
\nonumber\\
&&\nonumber\\
&=&
\int\limits_{4m^2}^\infty \frac{d\kappa^2}{\kappa^2}\cdot
\frac{\rho(\kappa^2)}{\kappa^2-p^2-i\epsilon}
=\frac{2}{p^2}
\left({\rm ArcSin}\left[\sqrt{\frac{p^2}{4\,m^2}}\right]\right)^2\,,
\nonumber
\end{eqnarray}
where 
$\rho(\kappa^2) = \ln\left\{\left(1+\sqrt{1-4m^2/\kappa^2}\right)/
          \left(1-\sqrt{1-4m^2/\kappa^2}\right)\right\}
$ 
is the spectral function.

Another interesting limiting case is
when one of the photons has large Euclidean momentum, for instance, 
$q_1^2=-Q^2$, $q_2^2=0$ and $p^2=0$. In the limit $Q^2\to \infty$
the integral Eq.(\ref{triangle_loc}) reduces to
\begin{eqnarray}\label{loc_asym}
I_{\rm L}(0,-Q^2,0) &\to&
\frac{1}{2Q^2}\int\limits_0^1 dx\,
\frac{\ln[1+x(1-x)Q^2/m^2]}{x(1-x)}
\to \frac{\ln^2(Q^2/m^2)}{2Q^2}.
\nonumber
\end{eqnarray}
Note that such an asymptotics is in contradiction to the QCD-prediction 
for the $\pi^0\gamma^\ast\gamma$ form factor of
$1/Q^2$~\cite{Brodsky_Lepage}.

In the following we turn to the evaluation of the
$\sigma\gamma\gamma$ diagram in the local limit using
dimensional regularization.
The defining integral of Eq.~(\ref{triangle}) can be written as 
\begin{eqnarray}\label{sigma} 
I^{\mu\nu; \, \sigma}_{\triangle,\,\rm L}(q_1,q_2) \, = \, 
F^{\sigma}_{\rm L}(p^2,q_1^2,q_2^2) \, b^{\mu\nu} \, + \,  
G^{\sigma}_{\rm L}(p^2,q_1^2,q_2^2) \, c^{\mu\nu}\,, 
\end{eqnarray}
with

\begin{eqnarray}\label{coeff}
F^{\sigma}_{\rm L}(p^2,q_1^2,q_2^2) 
&=& \frac{(I^{\sigma}_{\triangle, \, \rm L} \cdot b) \, c^2 -
              (I^{\sigma}_{\triangle, \, \rm L} \cdot c) \, (b \cdot c)}
             {b^2 c^2 - (b \cdot c)^2} \,,
\nonumber\\
&&\\
G^{\sigma}_{\rm L}(p^2,q_1^2,q_2^2)
&=& \frac{(I^{\sigma}_{\triangle, \, \rm L}\cdot c) \, b^2 -
              (I^{\sigma}_{\triangle,\,\rm L}\cdot b)\,(b\cdot c)}
             {b^2 c^2 - (b\cdot c)^2}\,,
\nonumber
\end{eqnarray}
and where the dot-product refers to the contraction of the Lorentz indices.

Using the $\alpha$-parametrization we have

\begin{eqnarray}\label{coeff_alp1}
F^\sigma_{\rm L}(p^2,q_1^2,q_2^2) &=& -\,m\,\int\!d^3\alpha \, 
\delta\left(1-\sum\limits_{i=1}^3\alpha_i\right)
\frac{1-4\,\alpha_1\,\alpha_2}{m^2-D_0}\,,
\nonumber\\
&&\\
G^\sigma_{\rm L}(p^2,q_1^2,q_2^2) &=& \frac{m}{q_1^2\,q_2^2}
\int\!d^3\alpha\,\delta\left(1-\sum\limits_{i=1}^3\alpha_i\right)
\frac{\alpha_1(1-2\,\alpha_1)\,q_1^2+\alpha_2(1-2\,\alpha_2)
\,q_2^2}{m^2-D_0}\,. \label{coeff_alp2}\nonumber
\end{eqnarray}

Again, one integration in Eq. (\ref{coeff_alp1}) can be performed 
analytically (the resulting expressions (\ref{coeff_res1}) and 
(\ref{coeff_res2}) are indicated in Appendix~C). 
For the case $q_1^2=q^2_2=0$ we get

\begin{eqnarray}
F^\sigma_{\rm L}(p^2,0,0) &=& -\,\frac{m}{4}\,\int\limits_0^1\! dv\,
\ln\left(\frac{1+\sqrt{1-v}}{1-\sqrt{1-v}}\right)
\frac{1-v}{m^2-v\cdot p^2/4}\,,
\nonumber\\
&&\\
G^\sigma_{\rm L}(p^2,0,0) &=& \frac{2\,m}{4\,m^2-p^2}\,\int\limits_0^1\! dv\,
\left\{\frac{1}{4}(1+v)\,
\ln\left(\frac{1+\sqrt{1-v}}{1-\sqrt{1-v}}\right)
-\sqrt{1-v}\right\}
\frac{1-v}{m^2-v\cdot p^2/4}\, .
\nonumber
\end{eqnarray}

\subsubsection{The nonlocal $\pi^0\gamma\gamma$ 
and $\sigma\gamma\gamma$ diagrams}

For a nontrivial, that is nonlocal, meson-quark vertex the evaluation of 
the relevant Feynman diagrams is based on a method outlined in \cite{Model}.
This technique was originally developed to derive a representation for 
the $\pi^0\gamma\gamma$ diagram (Fig.5a) with a dressed pion-quark
vertex, which is described by an arbitrary function $\tilde\Phi(-k^2)$ 
decreasing rapidly in the Euclidean region. It also preserves 
translational and gauge invariance in a manifest way. 

In our model the $\pi^0\gamma\gamma$ form factor is given by 
\begin{eqnarray}
F_{\pi\gamma\gamma}(p^2,q_1^2,q_2^2) &=& \frac{g}{2\pi^2\sqrt{2} } \, m \, 
I_{\rm NL}(p^2,q_1^2,q_2^2)
\end{eqnarray}
where $I_{\rm NL}$ is the nonlocal (NL) quark-loop integral: 
\begin{eqnarray}\label{pion-nonloc}
I_{\rm NL}(p^2,q_1^2,q_2^2) &=&
\int\frac{d^4k}{4\pi^2 i}\frac{\tilde\Phi(-k^2)}
{\left[m^2-(k+p/2)^2\right]
 \left[m^2-(k-p/2)^2\right]\left[m^2-(k+q/2)^2\right]}
\nonumber\\
&&\\
&=&
\int\limits_0^\infty\! dt
\left(\frac{t}{1+t}\right)^2
\int\!d^3\alpha \, \delta\left(1-\sum\limits_{i=1}^3\alpha_i\right)
\cdot\left\{-\tilde\Phi^\prime(z)\right\} 
\nonumber
\end{eqnarray}
with  $q=q_2-q_1$. 

The argument $z$ appearing in the vertex function is written in the form
\begin{eqnarray}\label{argument}
z&=&\frac{t^2}{1+t}\,D+\frac{t}{1+t}\,\Delta\,,
\nonumber\\
D&=&m^2-\alpha_1\alpha_2\,p^2-\alpha_1\alpha_3\,q_1^2-
\alpha_2\alpha_3\,q_2^2 \equiv m^2-D_0\,,
\nonumber\\
\Delta&=& m^2-\frac{p^2}{4}+\frac{\alpha_3}{2}\,(p^2-q_1^2-q_2^2)\,.
\nonumber
\end{eqnarray}
We choose to define the functional dependence of the vertex function
as $\tilde\Phi(-k^2)$, hence after transition to the Euclidean region
($k^0\to ik_4$ or $k^2\to -k^2_E$) the argument changes sign. This convention
allows a consistent choice of the functional form for $\tilde\Phi(k^2_E)$
in the Euclidean region, where we finally perform the integrations.
For example, particular forms of the vertex function are
$\tilde\Phi(k^2_E)=\exp(-k^2_E/\Lambda^2)$ (Gaussian  
vertex function) or $\tilde\Phi(k^2_E)=1/(1+k^2_E/\Lambda^2)^n$ 
(pole vertex function), where $1/\Lambda$ characterizes the size of the meson.
Obviously, the limiting local case $I_{\rm NL} \to I_{\rm L}$ can be recovered
for $\Lambda\to\infty$. 
This limit also serves as a check for the numerical calculations used in the
final step of the evaluation. The following analytical results are obtained
for an arbitrary form of $\tilde\Phi(-k^2)$. Only when turning to the final
numerical calculation, a specific form of $\tilde\Phi(k^2_E)$ 
(Gaussian vertex function) will be used.

Again, one integration in Eq.~(\ref{pion-nonloc}) is performed by
linearization of the argument $z$ with respect to one of the $\alpha$ 
parameters as it was done in the local case. The resulting expressions are 
indicated in Appendix D, as given by Eqs. (\ref{pi_NL1}) and (\ref{pi_NL2}).
For the study of the dependence of the $\pi\gamma\gamma$ form factor on the
photon virtualities in the Euclidean region it is useful to introduce the 
variables: $q_1^2=-(1+\omega)\,Q^2/2$ and $q_2^2=-(1-\omega)\,Q^2/2$ where
$Q^2$ and $\omega$ are the total virtuality and the asymmetry, 
respectively\footnote{Another possibility is to work in terms of the average  
virtuality. In this case the variable $Q^2$ should be rescaled as 
$Q^2 \to 2 Q^2$ in all further formulas.}.
With this convention we define the pion form factor as
\begin{eqnarray}
F_{\pi\gamma^\ast\gamma^\ast}(Q^2,\omega) \, \doteq \,  
F_{\pi\gamma\gamma}\biggl(M_\pi^2, - (1+\omega)\frac{Q^2}{2}, 
- (1-\omega)\frac{Q^2}{2} \biggr)   \, .
\end{eqnarray}  
The particular choice $\omega = \pm 1$ corresponds to the physically 
interesting case where one of the photons is virtual and the other is on 
its mass-shell. The corresponding transition form factor
$F_{\pi\gamma\gamma^\ast}(Q^2) \, \doteq \, 
F_{\pi\gamma\gamma}(M_\pi^2,-Q^2,0)$ 
has recently been measured by the CLEO 
Collaboration~\cite{Gronberg} for momentum transfers in the range from 
1.5 to 9 GeV$^2$. A discussion of previous experiments can also be found 
in Ref.~\cite{Gronberg}. 
The CLEO data confirmed the predictions of perturbative QCD 
(pQCD) \cite{Brodsky_Lepage} for the asymptotic behavior 
of $F_{\pi\gamma\gamma^\ast}(Q^2)$ at large $Q^2$: 
\begin{eqnarray}
F_{\pi\gamma\gamma^\ast}(Q^2) \, = \, \frac{2F_\pi}{Q^2} 
\, + \, O\biggl(\frac{1}{Q^4}\biggr)   
\end{eqnarray}  
where $F_\pi = f_\pi/\sqrt{2}\,$. 

A detailed analysis of the different QCD approaches to the 
$F_{\pi\gamma\gamma^\ast}(Q^2)$ form factor has recently been done 
in \cite{Bakulev}. Unfortunately, a straightforward calculation 
of $F_{\pi\gamma\gamma^\ast}(Q^2)$ in the context of QCD is either not
fully possible, 
because the operator product expansion (OPE) fails at small $Q^2$, or 
contains unknown parameters like the twist-four scale parameter 
$\delta^2$ related to the gluon condensate \cite{Chernyak,Novikov}.
Therefore, an analysis of $F_{\pi\gamma\gamma^\ast}$
in our QCD-motivated approach seems to be quite reasonable.
Information about the $H\to\gamma\gamma$ form factors for non-trivial photon 
virtualities is relevant to forthcoming experiments on production of 
pseudoscalar mesons in $\gamma^\ast\gamma^\ast$ collisions. Also, there 
is the project to search for a light scalar meson in very peripheral heavy 
ion collisions (for a recent review see Ref.~\cite{Baur}).

For the analysis of $F_{\pi\gamma^\ast\gamma^\ast}$ we perform a systematic
expansion in powers of $1/Q^2$ including leading order (LO) and
next-to-leading order (NLO) terms with 
\begin{eqnarray}\label{pigg_exp}
F_{\pi\gamma^\ast\gamma^\ast}(Q^2,\omega) \, = \, 2 \, F_\pi \, 
\biggl\{ \frac{J_\pi^{\rm LO}(\omega)}{Q^2} 
\, + \, \frac{J_\pi^{\rm NLO}(\omega)}{Q^4} \, + \, 
O\biggl(\frac{1}{Q^6}\biggr) \, \biggl\} \, . 
\end{eqnarray}   
The expansion coefficients $J_\pi^{\rm LO}(\omega)$ and 
$J_\pi^{\rm NLO}(\omega)$ are given in our approach by 
\begin{eqnarray}\label{jpi_exp}
J_\pi^{\rm LO}(\omega) \, = \,  \frac{2}{3} \, 
\frac{R_\pi^{\rm LO}(\omega)}{R_\pi^{\rm LO}(0)}\,, 
\hspace*{1cm}  
J_\pi^{\rm NLO}(\omega) \, = \, \frac{2}{3} \, 
\frac{R_\pi^{\rm NLO}(\omega)}{R_\pi^{\rm LO}(0)}\,. 
\end{eqnarray}
The structure integrals $R_\pi^{\rm LO}(\omega)$ and 
$R_\pi^{\rm NLO}(\omega)$ are derived in Appendix D  
applying the chiral limit with $M_\pi^2 = 0$ and are written as  
\begin{eqnarray}\label{R_pi}  
R_\pi^{\rm LO}(\omega) \, = \, \frac{1}{\omega} \, 
\int\limits_0^\infty\! \frac{dt}{t+1} \, \tilde\Phi(m^2 t) \,\,  
{\rm Ln}_-(t,\omega) \,,\hspace*{.75cm} 
R_\pi^{\rm NLO}(\omega) \,=\, - \, 4\,m^2 \, \int\limits_0^\infty\! dt \, 
\tilde\Phi(m^2 t) \,  D(t,\omega)\, , 
\end{eqnarray}   
where 
\begin{eqnarray}
{\rm Ln}_\pm(t,\omega) &\doteq& 
\ln\left[\frac{1+t\,(1+\omega)}{1+t}\right]
\, \pm \, \ln\left[\frac{1+t\,(1-\omega)}{1+t}\right]\,, \nonumber\\
D(t,\omega)&\doteq& \frac{t}{[1+t\,(1+\omega)]\,[1+t\,(1-\omega)]}\, . 
\nonumber
\end{eqnarray}
In the derivation of Eqs.~(\ref{pigg_exp}) and (\ref{jpi_exp}) we use 
the identity
\begin{eqnarray}\label{g_math_Fpi}  
F_\pi \, = \, \frac{3 \, g}{8 \,\pi^2 \sqrt{2}} \,\, R_\pi^{\rm LO}(0)\, ,
\end{eqnarray} 
relating $F_\pi$ and the meson-quark coupling constant $g$. 
From the expressions of Eq. (\ref{R_pi}) it can be readily seen
that the expansion coefficients contain only even powers of $\omega$, 
which is a consequence of Bose-Einstein symmetry and charge 
conjugation invariance. 

At this stage of the development we indicate a first comparison of our 
results for the expansion coefficients calculated at $\omega = 1$ and 
$\omega = 0$ to the ones predicted by 
pQCD~\cite{Brodsky_Lepage,Chernyak,Novikov}. A summary of previous model 
results for $J_\pi^{\rm LO}(1)$ and $J_\pi^{\rm LO}(0)$ can be found 
in Ref.~\cite{Tandy}. First, we consider the experimentally accessible 
case $w = \pm 1$, where one of the photons is 
real and and the other one is virtual. Our numerical result for 
$J_\pi^{\rm LO}(1) = 0.998 \simeq 1$ is in very good agreement with 
the pQCD prediction of $J_\pi^{\rm LO}(1) \equiv 1$ \cite{Brodsky_Lepage}. 
The deviation of our value for $J_\pi^{\rm LO}(1)$ from 1 is explained 
by the use of simplest form of quark propagator and meson-quark vertex 
form factor.

In the relativistic quark model the NLO coefficient $J_\pi^{\rm NLO}(1)$  
is proportional to the constituent quark 
mass squared and results in $J_\pi^{\rm NLO}(1) = - 0.37$ GeV$^2$. 
The pQCD approach predicts~\cite{Bakulev}  
\begin{eqnarray}
J_\pi^{\rm NLO}(1) \, = \, - \, \frac{80}{27} \, \delta^2 \, ,  
\end{eqnarray}
where $\delta^2$ is the twist-four scale parameter related to the gluon 
condensate~\cite{Chernyak,Novikov}. Originally, this quantity was estimated 
to have a value of $\delta^2 = (0.2 \pm 0.02)$ GeV$^2$ using the QCD sum 
rule approach~\cite{Novikov}. A recent evaluation \cite{Bakulev} results in: 
$\delta^2 = (0.19 \pm 0.02)$ GeV$^2$. By taking into account the value of 
Ref.~\cite{Novikov} the pQCD prediction is 
\begin{eqnarray}
J_\pi^{\rm NLO}(1) \, = \, -  \, (0.59 \pm 0.06) \,\,\,\, {\rm GeV}^2 \,. 
\end{eqnarray}
Now we turn to the case $\omega = 0$. We exactly reproduce the LO coefficient 
predicted by pQCD~\cite{Novikov}: $J_\pi^{\rm LO}(0) = 2/3$. For the NLO 
coefficient our result is 
\begin{eqnarray}
J_\pi^{\rm NLO}(0) \, = \, - \, \frac{4}{3} \, m^2 \, = \, - \, 0.074 
\,\,\,\, {\rm at} \,\,\,\, m = 235 \,\,\,\, {\rm MeV}\,. 
\end{eqnarray}
The result of pQCD is again proportional to $\delta^2$~\cite{Novikov}: 
\begin{eqnarray}
J_\pi^{\rm NLO}(0) \, = \, -  \, \frac{32}{27} \, \delta^2 \, = \, 
-(0.24 \pm 0.02) \,\,\,\, {\rm GeV}^2 \,. 
\end{eqnarray} 
From this comparison we conclude that our prediction for the LO expansion 
coefficients $J_\pi^{\rm LO}(1)$ and $J_\pi^{\rm LO}(0)$ are in perfect 
agreement with the expectation of pQCD. The NLO coefficients are rather 
sensitive to the choice of the constituent quark mass. In particular, when 
using $m = 395$ MeV in the calculation of $J_\pi^{\rm NLO}(1)$ and 
$m = 420$ MeV in the case of $J_\pi^{\rm NLO}(0)$ the central values of 
the NLO coefficients predicted by pQCD can be fitted. 

For the process $\sigma\to\gamma\gamma$ we have to evaluate, in addition to 
the triangle diagram (Fig.5a), the bubble (Figs.5b and 5c) and tadpole 
diagrams (Fig.5d), which arise from gauging the nonlocal $\sigma q \bar q$ 
interaction Lagrangian. Each particular diagram is not gauge invariant by 
itself, but the total sum fulfills the gauge invariance requirement.
To simplify the calculation we split the contribution of each diagram into
a part which is gauge invariant and one which is not.
This separation can be achieved in the following manner.
For the $\gamma$-matrices and vectors
with open Lorentz indices $\mu$ and $\nu$ we use the representation:

\begin{eqnarray}\label{split}
\gamma^\mu &=& \gamma^\mu_{\perp; \, q_1} \, + \, q_1^\mu \,  
\frac{\not\! q_1}{q_1^2}\,,  \hspace*{1cm}
v^\mu \, = \, v^\mu_{\perp; \, q_1} \, + \, q_1^\mu \, \frac{v q_1}{q_1^2}\,,
\nonumber\\
&&\\
\gamma^\nu &=& \gamma^\nu_{\perp; \, q_2} \, + \, q_2^\nu \, 
\frac{\not\! q_2}{q_2^2}\,, \hspace*{1cm}
v^\nu \, = \, v^\nu_{\perp; \, q_2} \, + \, q_2^\nu \, \frac{v q_2}{q_2^2}\,,
\nonumber 
\end{eqnarray}
such that $\gamma^\mu_{\perp; \, q_1} \,  (q_1)_\mu \, = \, 
v^\mu_{\perp; \, q_1} \, (q_1)_\mu=0$ and     
$\gamma^\nu_{\perp; \, q_2} \, (q_2)_\nu \, = \, v^\nu_{\perp; \, q_2} 
\,  (q_2)_\nu \, = \, 0$.
Expressions for diagrams containing only $\perp$-values are gauge invariant
separately.
It is easy to show that the remaining terms, which are not gauge invariant,
cancel each other in total.

The gauge-invariant parts of the nonlocal 
$\sigma\gamma\gamma$-triangle $(\triangle)$, 
bubble (bub) and tadpole (tad) diagrams are given by
\begin{eqnarray}
I^{\mu\nu}_{\triangle_\perp}(q_1,q_2) 
&=& \int\!\frac{d^4 k}{4i\pi^2}\,\tilde\Phi(-k^2)
{\rm tr}\bigg\{\gamma^\mu_{\perp; \, q_1} S(\not\! k + \frac{\not\! p}{2})
S\biggl(\not\! k - \frac{\not\! p}{2}\biggr) 
\gamma^\nu_{\perp; \, q_2} S\biggl(\not\! k + \frac{\not\! q}{2}\biggr)\bigg\}
\nonumber\\
&&\nonumber\\
&=& F^{\sigma}_{\triangle_\perp}(p^2,q_1^2,q_2^2) \, b^{\mu\nu}+
    G^{\sigma}_{\triangle_\perp}(p^2,q_1^2,q_2^2) \, c^{\mu\nu}\,, 
\label{sigma-trn}\\  
&&\nonumber\\
I^{\mu\nu}_{\rm bub_\perp}(q_1,q_2)
&=& -\,\int\!\frac{d^4 k}{4\pi^2 i}\int\limits_0^1\!d\tau\,
\tilde\Phi^\prime(- x(0,q_2)) 
\left \{k^\nu_{\perp; \, q_2} \,
{\rm tr}\left[\gamma^\mu_{\perp; \, q_1} S(\not\! k+\not\! q_1/2)
S(\not\! k-\not\! q_1/2)\right]\,
\right\}\nonumber\\
&&\nonumber\\
&+&(q_1\leftrightarrow q_2,\,\mu\leftrightarrow \nu)
\, = \,  F^{\sigma}_{\rm bub_\perp}(p^2,q_1^2,q_2^2) \, b^{\mu\nu} +
         G^{\sigma}_{\rm bub_\perp}(p^2,q_1^2,q_2^2) \, c^{\mu\nu}\,, 
\label{sigma-bub}\\
&&\nonumber\\
I^{\mu\nu}_{\rm tad_\perp}(q_1,q_2) 
&=& \int\!\frac{d^4 k}{4\pi^2 i}{\rm tr}\left[S(\not\! k)\right] 
\int\limits_0^1\!d\tau\, 
\biggl[ - \, \frac{c^{\mu\nu}}{4\,q_1^2\,q_2^2}\cdot 
\left\{ \tilde\Phi^\prime(- x(0,p) ) 
+ \tilde\Phi^\prime(- x(0,q) ) \right\}
\nonumber\\
&&\nonumber\\
&+& \tau \, \int\limits_0^1\!dl\, 
\biggl(k+\frac{q_2}{2}\biggr)^\mu_{\perp; \, q_1} \, 
k^\nu_{\perp; \, q_2} \, 
\biggl\{\tilde\Phi^{\prime\prime}(- x(q_1,q_2)) + 
\tilde\Phi^{\prime\prime}(- x(- q_1,q_2))\biggr\}\biggr]\nonumber\\
&&\nonumber\\
&+&(q_1\leftrightarrow q_2,\,\mu\leftrightarrow \nu)
\, = \, F^{\sigma}_{\rm tad_\perp}(p^2,q_1^2,q_2^2) \, b^{\mu\nu}+
        G^{\sigma}_{\rm tad_\perp}(p^2,q_1^2,q_2^2) \, c^{\mu\nu}\,,  
\label{sigma-tad} 
\end{eqnarray}
where
\begin{eqnarray*}
x(q_1,q_2) \, = \,  k^2+k\tau(l\,q_1+q_2)
            +\frac{\tau}{4}(l\,q_1^2+2\,l\,q_1q_2+q_2^2)\, . 
\end{eqnarray*}
Applying the $\alpha$-parametrization for the gauge-invariant part of 
the triangle diagram gives 
\begin{eqnarray}
F^{\sigma}_{\triangle_\perp} &=& \!-m \int\limits_0^\infty\! dt
\frac{t^3}{(1+t)^4}
\int\! d^3\alpha \, \delta\left(1-\sum\limits_{i=1}^3 \alpha_i\right)\,
\left\{-\tilde\Phi^{\, \prime}(z)\right\}\,
\left[t\,(1-4\,\alpha_1\,\alpha_2)+2\,\alpha_3\right], \label{coeff_trn_alp1}
\\
&&\nonumber\\
&&\nonumber\\
G^{\sigma}_{\triangle_\perp} &=& -\,\frac{m}{q_1^2\,q_2^2}
\int\limits_0^\infty\! dt\,\frac{t^2}{(1+t)^2}
\int d^3\alpha \, \delta\left(1-\sum\limits_{i=1}^3 \alpha_i\right)\,
\left\{-\tilde\Phi^{\, \prime}(z)\right\}\,
\label{coeff_trn_alp2}\\
&&\nonumber\\
&\times&
\left\{
- \, m^2 + \frac{(1+2\,\alpha_1\,t)(1+2\,\alpha_2\,t)}{4 (1+t)^2}\,p^2
+\frac{t\,(\alpha_1-\alpha_2)}{2 (1+t)^2}
\left((1+2\,\alpha_1\,t)\,q_1^2-(1+2\,\alpha_2\,t)\,q_2^2\right)
\right\}\,.
\nonumber
\end{eqnarray}
Again, using the t'Hooft-Veltman technique one integration  
in Eqs.~(\ref{coeff_trn_alp1}) and (\ref{coeff_trn_alp2}) can be performed.  
The resulting expressions are indicated in Eqs. (\ref{coeff_trn1})  
and (\ref{coeff_trn2}) of Appendix~E. The analytical results for the form 
factors $F(G)^{\sigma}_{\rm bub_\perp}$ and $F(G)^{\sigma}_{\rm tad_\perp}$ 
are given in Appendix~F (Eqs. (\ref{coeff_bub1})-(\ref{coeff_tad2})). 
 
In the local limit, that is $\Lambda\to\infty$, we obtain
$$
\def\arraystretch{1.5}
\begin{array}{lll}
F^{\sigma}_{\triangle_\perp} \to F^{\sigma}_{\rm L}\,,
&\hspace*{1cm}
F^{\sigma}_{\rm bub_\perp} \to 0\,,  
&\hspace*{1cm} 
F^{\sigma}_{\rm tad_\perp} \to 0\,,  
\\
G^{\sigma}_{\triangle_\perp} \to G^{\sigma}_{\rm L} 
\, + \, \frac{\displaystyle{m}}{\displaystyle{2\,q_1^2\,q_2^2}}\,,
&\hspace*{1cm}
G^{\sigma}_{\rm bub_\perp} \to  - \, 
\frac{\displaystyle{m}}{\displaystyle{q_1^2\,q_2^2}}\,,
&\hspace*{1cm}
G^{\sigma}_{\rm tad_\perp} \to \, 
\frac{\displaystyle{m}}{\displaystyle{2\,q_1^2\,q_2^2}}\,. 
\end{array}
$$
We therefore recover the local $\sigma\gamma\gamma$ form factors: 
\begin{eqnarray*}
F^{\sigma}_{\rm NL} &=& F^{\sigma}_{\triangle_\perp} \, + \, 
F^{\sigma}_{\rm bub_\perp} \, + \, F^{\sigma}_{\rm tad_\perp}   
\, \to \, F^{\sigma}_{\rm L}\,,\\
G^{\sigma}_{\rm NL} &=& G^{\sigma}_{\triangle_\perp} \, + \, 
G^{\sigma}_{\rm bub_\perp} \, + \, G^{\sigma}_{\rm tad_\perp}   
\, \to \, G^{\sigma}_{\rm L}\, ,
\end{eqnarray*}
which were already indicated in Eq. (\ref{coeff_alp1}).

The form factors $F_{\sigma\gamma\gamma}$ and 
$G_{\sigma\gamma\gamma}$ are expressed in terms of 
the functions introduced above as 
\begin{eqnarray}
F_{\sigma\gamma\gamma}(p^2,q_1^2,q_2^2) &=& 
- \, \frac{5g}{6\pi^2\sqrt{2} } \, 
\biggl[ F^{\sigma}_{\triangle_\perp}(p^2,q_1^2,q_2^2) 
\, + \, F^{\sigma}_{\rm bub_\perp}(p^2,q_1^2,q_2^2) 
\, + \, F^{\sigma}_{\rm tad_\perp}(p^2,q_1^2,q_2^2) \biggr]\,,\\ 
G_{\sigma\gamma\gamma}(p^2,q_1^2,q_2^2) &=& 
- \, \frac{5g}{6\pi^2\sqrt{2} } \, 
\biggl[ G^{\sigma}_{\triangle_\perp}(p^2,q_1^2,q_2^2) 
\, + \, G^{\sigma}_{\rm bub_\perp}(p^2,q_1^2,q_2^2) 
\, + \, G^{\sigma}_{\rm tad_\perp}(p^2,q_1^2,q_2^2) \biggr]\,. 
\end{eqnarray}
In analogy to the case of the pion we define the $\sigma\gamma\gamma$ 
form factors using the $Q^2$ and $\omega$ variables accordingly: 
\begin{eqnarray}
F_{\sigma\gamma^\ast\gamma^\ast}(Q^2,\omega) &\doteq& 
F_{\sigma\gamma\gamma}\biggl(M_\sigma^2, - (1+\omega)\frac{Q^2}{2}, 
- (1-\omega)\frac{Q^2}{2} \biggr)\,, \\
G_{\sigma\gamma^\ast\gamma^\ast}(Q^2,\omega) &\doteq& 
G_{\sigma\gamma\gamma}\biggl(M_\sigma^2, - (1+\omega)\frac{Q^2}{2}, 
- (1-\omega)\frac{Q^2}{2} \biggr)\,.   
\end{eqnarray}
The final numerical analysis indicates that the additional nonlocal
diagrams (bubble and tadpole) are 
significantly suppressed. For simplicity below we discuss the power expansion 
for $F_{\sigma\gamma^\ast\gamma^\ast}(Q^2,\omega)$ and 
$G_{\sigma\gamma^\ast\gamma^\ast}(Q^2,\omega)$ in the limit 
$M_\sigma^2 = 0\,$, where only the gauge-invariant part of the triangle
diagram is included. 
In the numerical analysis (Sec.~V) we will take into account all diagrams 
and use the value of $M_\sigma = 385.4$ MeV as predicted by our approach.
For the power expansion of the $\sigma\gamma\gamma$ form factor we pull out,
with the help of Eq.~(\ref{g_math_Fpi}), a common scaling factor $2 F_\pi$:
\begin{eqnarray}
F_{\sigma\gamma^\ast\gamma^\ast}(Q^2,\omega) &=& 2F_\pi \, 
\biggl\{ \frac{J_\sigma^{\rm LO}(\omega)}{Q^2} 
\, + \, \frac{J_\sigma^{\rm NLO}(\omega)}{Q^4} \, + \, 
O\biggl(\frac{1}{Q^6}\biggr) \, \biggl\}\,,\nonumber\\
&&\\
G_{\sigma\gamma^\ast\gamma^\ast}(Q^2,\omega) &=& 2F_\pi \, 
\biggl\{ \frac{K_\sigma^{\rm LO}(\omega)}{Q^4} 
\, + \, \frac{K_\sigma^{\rm NLO}(\omega)}{Q^6} \, + \, 
O\biggl(\frac{1}{Q^8}\biggr) \, \biggl\}\,. \nonumber 
\end{eqnarray}   
The expansion coefficients  $J_\sigma^{\rm LO}(\omega)$, 
$J_\sigma^{\rm NLO}(\omega)$, $K_\sigma^{\rm LO}(\omega)$ and 
$K_\sigma^{\rm NLO}(\omega)$ are derived in the limit indicated above as 
\begin{eqnarray}
J_\sigma^{\rm LO}(\omega) &=& \frac{5}{3 \, \omega^2} 
\, \biggl\{ \, J_\pi^{\rm LO}(\omega) \, -  \, \frac{2}{3} \, \biggr\}\,, 
\hspace*{.7cm}  
J_\sigma^{\rm NLO}(\omega) \, = \, \frac{10}{9} 
\frac{R_{\sigma 1}^{\rm NLO}(\omega)}{R_\pi^{\rm LO}(0)}\,,\nonumber\\
&&\\
K_\sigma^{\rm LO}(\omega) & = & - 2 J_\sigma^{\rm LO}(\omega)\,, 
\hspace*{2.6cm}  K_\sigma^{\rm NLO}(\omega) \, = \, 
- \, 2 \, J_\sigma^{\rm NLO}(\omega) \, - \, \frac{10}{9} \, 
\frac{R_{\sigma 2}^{\rm NLO}(\omega)}{R_\pi^{\rm LO}(0)}\,,\nonumber
\end{eqnarray}
where 
\begin{eqnarray}\label{R_sigma}  
R_{\sigma 1}^{\rm NLO}(\omega) &=&  - \, \frac{4 \, m^2}{\omega^2} \, 
\int\limits_0^\infty\! dt \, 
\tilde\Phi(m^2 t) \, 
\biggl\{ \frac{t}{1+t} \, [ \, 1 \, +  \, D(t,\omega) \, ] 
\, - \, \frac{3-\omega^2}{\omega^2} \,\, {\rm Ln}_+(t,\omega) \,\nonumber\\  
&-& \, \frac{1+2\,t}{(1+t) \, \omega} \,\, {\rm Ln}_-(t,\omega)\biggr\}\,,\\
&&\nonumber\\
R_{\sigma 2}^{\rm NLO}(\omega) &=& \frac{8 m^2}{(1 - \omega^2)} 
\int\limits_0^\infty\! dt \, \tilde\Phi( m^2 t) \, 
\biggl\{ \, \frac{t}{t+1} \, - \, 
\frac{2-\omega^2}{\omega^2} \,\, {\rm Ln}_+(t,\omega) \,  
- \, \frac{1}{\omega} \,\, {\rm Ln}_-(t,\omega) \,  
\biggr\}\,.\nonumber
\end{eqnarray}   
Final numerical values for the expansion coefficients are: 
\begin{eqnarray}
&&J_\sigma^{\rm LO}(1) = - \frac{K_\sigma^{\rm LO}(1)}{2} = 0.54\,, 
\hspace*{1cm} 
J_\sigma^{\rm LO}(0) = - \frac{K_\sigma^{\rm LO}(0)}{2} = 0.23\,, 
\nonumber\\[2mm]
&&J_\sigma^{\rm NLO}(1) = 1.18 \,\,\, {\rm GeV}^2\,, \hspace*{1.9cm} 
J_\sigma^{\rm NLO}(0) = 0.87 \,\,\, {\rm GeV}^2\,, \\
&&\nonumber\\
&&K_\sigma^{\rm NLO}(1) = 4.61 \,\,\, {\rm GeV}^2\,, \hspace*{1.8cm} 
K_\sigma^{\rm NLO}(0) = - 0.09 \,\,\, {\rm GeV}^2\,. \nonumber 
\end{eqnarray}

\section{The sigma meson in strong and weak decays} 

In this section we derive the matrix elements describing the strong 
decay $\sigma\to\pi\pi$ and the nonleptonic decays
$D\rightarrow\sigma\pi$ and $B\rightarrow\sigma\pi$.  

\subsection{Strong decay $\sigma\to\pi\pi$}\label{strongdecay}
The strong form factor $G_{\sigma\pi\pi}(p^2,q_1^2,q_2^2)$ 
related to the transition $\sigma\to \pi\pi$ is defined in our approach 
as 
\begin{eqnarray}\label{gsigmapipi}
G_{\sigma\pi\pi}(p^2,q_1^2,q_2^2) &=& \frac{3 g^3}{2\pi^2 \sqrt{2}} 
\int \frac{d^4k}{4\pi^2 i} \, \tilde\Phi(- k^2) \, 
\tilde\Phi\biggl(- \biggl[k + \frac{p_2}{2}\biggr]^2 \biggr) \, 
\tilde\Phi\biggl(- \biggl[k - \frac{p_1}{2}\biggr]^2 \biggr) \, \\
&&\nonumber\\
&\times&{\rm tr}\biggl\{ \gamma^5 S\biggl(k + \frac{p}{2}\biggr) 
S\biggl(k - \frac{p}{2}\biggr) \gamma^5 S\biggl(k + \frac{q}{2}\biggr) \, .
\biggr\}\nonumber
\end{eqnarray}
The $\sigma $ momentum is given by $p$, pion momenta are labeled by
$q_1$ and $q_2$ with $p = q_1 + q_2$ and  $q = q_2 - q_1$. The matrix element
of Eq. (\ref{gsigmapipi}) was already introduced in its generic form
by Eq. (\ref{Lambda_Trans}), when we discussed the model features.
The corresponding evaluation is summarized in Appendix A. 

Since the strong modes $\sigma\to\pi^+\pi^-$ and $\sigma\to\pi^0\pi^0$
dominate the $\sigma$ decays, the total width $\Gamma_\sigma$ is given by
\begin{eqnarray}
\Gamma_\sigma \, \simeq \, \Gamma(\sigma\to\pi^+\pi^-) \, 
+ \, \Gamma(\sigma\to\pi^0\pi^0) \, .
\end{eqnarray}
With the coupling constant
$g_{\sigma\pi\pi} \doteq G_{\sigma\pi\pi}(M_\sigma^2,M_\pi^2,M_\pi^2)$
the total decay width $\Gamma_\sigma$ can then be expressed as
\begin{equation}\label{dec_sigma} 
\Gamma_\sigma \, = \, \frac{3}{2} \, \Gamma(\sigma\rightarrow\pi^+\pi^-)
=\frac{3 \, g_{\sigma\pi\pi}^2}{32 \, \pi \, M_\sigma^3} \, \lambda^{1/2}
(M_\sigma^2,M_\pi^2,M_\pi^2),
\end{equation}
where $\lambda^{1/2} (M_\sigma^2,M_\pi^2,M_\pi^2)$ was already defined 
in Eq. (\ref{lambda_tri}). 
 
In our estimate for the $G_{\sigma\pi\pi}$ form factor and
the $\sigma$-meson width final state interaction is neglected.
An accurate analysis of the strong decay properties of the $\sigma$
meson should also include these effects.
We just refer to two Refs. \cite{Colangelo,Oller} where the $\sigma$-meson
was generated dynamically in the iso-scalar S-wave of $\pi\pi$
scattering in the context of chiral perturbation theory.
The pole identified with the light $\sigma$ meson occurs
at $E = M_\sigma - (i/2) \Gamma_\sigma$ with $M_\sigma \approx 500$ MeV
and $\Gamma_\sigma \approx $ 600 MeV \cite{Colangelo} and
400 MeV \cite{Oller}. Judging from this work, an inclusion of final state
interaction will possibly lead to a further increase of $\Gamma_\sigma$ as
compared to the value derived in our model.

\subsection{Weak decays $D\rightarrow\sigma\pi$ and $B\rightarrow\sigma\pi$} 

Now we turn to the discussion of the nonleptonic two-body transitions
$H\rightarrow\sigma\pi$ with $H = D$ or $B$. 
The effective interaction Lagrangian relevant for the nonleptonic 
two-body decays $D\to \sigma\pi$ and $B\to \sigma\pi$ is given  
by~\cite{Buchalla,Stech,Cheng,Beneke}: 
\begin{eqnarray}\label{Lagr_nl}
{\cal L}^{nl}_{\rm int} = 
- \frac{G_F}{\sqrt{2}} \biggl\{ V^{\ast}_{cd} V_{ud} \, ( \, 
 a_1^c  O_1^c \, + \,  a_2^c  O_2^c ) \,  
+ \, V^{\ast}_{ub} V_{ud} \, ( \, a_1^b O_1^b \, + 
\, a_2^b O_2^b \, ) \biggr\} \, + \, {\rm H.c.}  
\end{eqnarray}
where $G_F$ is the Fermi coupling constant and $V_{qq^\prime}$ are the 
matrix elements of the Cabibbo-Kabayashi-Maskawa quark-mixing matrix 
($V^{\ast}_{cd} \, V_{ud} \, = \, 0.217$ and 
$V^{\ast}_{ub} \, V_{ud} \, = \, 0.0036$ )~\cite{PDG}.
The four-quark operators $O_i^Q$ are defined as:
\begin{eqnarray}
& &O_1^c = (\bar d_i c_i)_{V-A} (\bar u_j d_j)_{V-A}, \,\,\, 
   O_2^c = (\bar u_i c_i)_{V-A} (\bar d_j d_j)_{V-A}, \,\,\, 
\\
& &O_1^b = (\bar u_i b_i)_{V-A} (\bar d_j u_j)_{V-A}, \,\,\, 
   O_2^b = (\bar d_i b_i)_{V-A} (\bar u_j u_j)_{V-A}, \,\,\, 
\end{eqnarray}
where $i,j$ are the color indices and label $V-A$ is a short-hand notation for
the $\gamma^\mu(1 - \gamma^5)$ Dirac structure. 
The couplings $a_i^{Q}$ are the combination of the 
Wilson coefficients including both the factorizable and 
the nonfactorizable effects. We use the following values for the       
effective couplings $a_1^{Q}$: $a_1^{c} = 1.274$~\cite{Cheng} and 
$a_1^{b} = 1.038$~\cite{Beneke}. 

The corresponding decay width $\Gamma(H\to\sigma\pi)$ is given by 
\begin{equation}\label{weakwidth}
\Gamma(H\rightarrow\sigma\pi)=\frac{g_{H\sigma\pi}^2}{16\pi M_H^3}
\lambda^{1/2}(M_{H}^2,M_\sigma^2,M_\pi^2),
\end{equation}
where $g_{H\sigma\pi}$ is the effective weak coupling constant. The constant 
$g_{H\sigma\pi}$ is equivalent to the expectation value of the effective 
Hamiltonian ${\cal H}_{eff}$ as derived from the nonleptonic Lagrangian 
(\ref{Lagr_nl})
\begin{equation} \label{eq_gds} 
g_{H\sigma\pi}=<\sigma\pi|{\cal H}_{eff}|H>=\frac{G_F}{\sqrt{2}} \, 
V_{cd(ub)}^\ast V_{ud} \, a_1^Q \, 
F_0^{H\sigma}(M_\pi^2) \, (M_H^2 - M_\sigma^2) \, f_\pi,
\end{equation}
where $F_0^{H\sigma}$ is the weak form factor describing the 
$H\to\sigma$ transition~\cite{Dib,Gatto}. 

In Refs.~\cite{Gatto,Deandrea} the authors included two 
contributions to the form factor $F_0^{H\sigma}$ : 
the direct diagram of Fig.6a and the resonance or polar diagram 
of Fig.6b with an intermediate axial meson $H(1^+)$. The contribution 
of the resonance diagram is sizeable and close to the contribution of 
the direct diagram. As result they overestimated the experimental result 
for $\Gamma(D^+\rightarrow\sigma\pi^+)$. Therefore, we restrict ourselves
to the consideration of the direct diagram (Fig.6a) only. 
We neglect also the suppressed "annihilation" diagram,
that is the $D \to W \to \sigma \, + \, \pi$
transition, since it involves form factors of 
light mesons at high momentum transfer ($q^2=M_{D}^2$)~\cite{Stech}. 
Note, that the diagram generated by operator $O_2^Q$ is vanishing 
because the corresponding matrix element is proportional to 
$<0| (\bar q q)_{V-A} |\sigma> = 0$. 

The contribution of the direct diagram (Fig.6a) to the form factor $F_0^{H\sigma}$ 
is given by 
\begin{eqnarray}\label{weakform}
F_0^{H\sigma}(q^2) &=& \frac{3 g_{H} g_\sigma}
{4 \pi^2 \sqrt{2} \, (M_H^2 - M_\sigma^2)} \, I_{H\sigma}(q^2)\,  
\end{eqnarray}
where $I_{H\sigma}$ is the structure integral: 
\begin{eqnarray} 
I_{H\sigma}(q^2) &=& \int \frac{d^4k}{4\pi^2i}  \, 
\tilde\Phi( - \biggl[k + \frac{p^{\, \prime}}{2}\biggr]^2) \, 
\tilde\Phi_H( - (k + \omega_{Qq} p)^2) \nonumber\\ 
&&\nonumber\\
&\times&{\rm tr}\left[S_Q(\not\! k + \not\! p) \, \gamma^5 \, 
S(\not\! k) \, S(\not\! k + \not\! p^{\, \prime}) \not\! q 
\gamma^5 \right]\,  
\end{eqnarray} 
with $\omega_{Qq} = m_q/(m_Q+m_q)$. 
Again, the analytical evaluation of the structure integral, before
the final numerical calculation is applied, is indicated in Appendix A. 

\section{Numerical analysis}

\subsection{Electromagnetic form factors}

Preliminary model results for the pion charge $F_\pi(Q^2)$ 
and $F_{\pi\gamma\gamma^\ast}(Q^2)$ form factors were
already presented in Ref. \cite{Model}. Here we extend our formalism to 
the case of the $\sigma\gamma\gamma$ form factors and we also perform 
a comprehensive analysis of the $H\to\gamma\gamma$ form factors. 

Recently, new and more accurate experimental results for the charged 
pion electromagnetic form factor $F_\pi(Q^2)$ were obtained by
the Jefferson Lab $F_\pi$ Collaboration \cite{Volmer}. These data for the
momentum transfer region of $Q^2 = 0.6 - 1.6$ GeV$^2$ were extracted
from an analysis of electro-production of pions on the nucleon.
The new results for the pion form factor lie somewhat higher
than the older Cornell data points \cite{Bebek}, but are consistent with 
a monopole parametrization fitted to elastic data at very low $Q^2$ and 
with the $1/Q^2$ scaling law~\cite{Matveev}.
In Fig.7 we show our results for 
$F_\pi(Q^2)$ in the region up to $Q^2 = 4$ GeV$^2$. For comparison we 
indicate the experimental data, recent results ($F_\pi$ 
Collaboration \cite{Volmer}) and previous ones (DESY~\cite{Brauel} and 
CERN NA7~\cite{Amendolia} Collaboration). We also indicate the predictions
of other theoretical approaches: QCD sum rules \cite{Nesterenko_Radyushkin}, 
light-cone quark model~\cite{Gardarelli}, NJL model with a 
separable $q \bar q$ interaction \cite{Gross} and QCD modeling 
approach based on solutions of 
the Dyson-Schwinger equations~\cite{Maris}. 
Our model predictions provide
a rather good description of the available data and are very close
to the QCD sum rule result \cite{Nesterenko_Radyushkin} including $O(\alpha_s)$
correction. They are also close to the results of the QCD motivated
approach~\cite{Maris}, which is based on a similar physical picture. 

In contrast to the work of Ref. \cite{Maris} we
use, roughly speaking, simple phenomenological prescriptions
for the quark propagator (a free propagator with an effective quark mass instead
of a confined one) and for the meson correlation function.
We also get a reasonable description 
of the pion charge radius with $r_{\pi} =  0.65$ fm, as obtained in our model.
Our result should be compared to the present
world average data of $r_{\pi} = (0.672 \pm 0.008)$ fm 
from PDG2002~\cite{PDG}, to the recent experimental result 
$r_{\pi} = (0.65 \pm 0.05 \pm 0.06)$ fm of the SELEX 
Collaboration~\cite{Eschrich} and to the prediction of
Ref.~\cite{Maris} with $r_{\pi} = 0.67$ fm. 

Next we discuss the numerical results for the $\pi^0 \to \gamma\gamma$ 
transition form factor. First we consider our results for the form factor 
$F_{\pi\gamma\gamma^\ast}(Q^2)$, which are given in Fig. 8. The data points
are taken from \cite{Behrend} (CELLO) and \cite{Gronberg} (CLEO). 
Other theoretical calculations include the hard scattering approach 
(HSA)~\cite{Jakob}, QCD sum rules~\cite{Radyushkin} and perturbative 
light-cone QCD~\cite{Cao}. Our curve for $Q^2 F_{\gamma^\ast\gamma\pi}(Q^2)$ 
is in good agreement with the data and approaches the Brodsky-Lepage limit 
for large $Q^2$. With the usual definition for the range of a form factor, 
that is
\begin{eqnarray}\label{rad_H}
<r^2_{H\gamma}> = - 6 \frac{dF_{H\gamma\gamma^\ast}(Q^2)}{dQ^2}
\Bigg|_{\displaystyle{Q^2 = 0}} \,\,\,\,
{\rm with} \,\,\,\, H = \pi, \sigma ,
\end{eqnarray}
we obtain for the radius of $F_{\pi\gamma\gamma^\ast}(Q^2)$:
\begin{equation}
<r^2_{\pi\gamma}> = 0.44 \,\,{\rm fm}^2 \, .
\end{equation}
Our result confirms
the monopole-type approximation of the CLEO data~\cite{Gronberg} and 
is very close to the CELLO measurement~\cite{Behrend} of
$<r^2_{\pi\gamma}> = 0.42 \pm 0.04$ fm$^2$. Again, our model prediction
is close to the result of a similar theoretical approach with
$<r^2_{\pi\gamma}> = 0.39 \pm 0.04$~fm$^2$~\cite{Tandy}.
Fig.9 contains our results 
for the $F_{\pi\gamma^\ast\gamma^\ast}(Q^2,\omega)$ form factor for
different values of the asymmetry parameter $\omega = 1, \, 3/4, \, 1/2$ 
and $0$. An increase in $\omega$ leads to a rise of 
$F_{\pi\gamma^\ast\gamma^\ast}(Q^2,\omega)$
for large $Q^2$. 
Finally, for the coupling constant 
$g_{\pi\gamma\gamma}$ and the decay width $\Gamma_{\pi\gamma\gamma}$ 
our approach gives: 
\begin{eqnarray}
g_{\pi\gamma\gamma} \, = \, 0.263 \,\,\, {\rm GeV}^{-1} 
\hspace*{1cm} {\rm and} \hspace*{1cm} 
\Gamma_{\pi\gamma\gamma} \, = \, 7.15 \,\,\, {\rm eV} \,  
\end{eqnarray} 
which is close to the data~\cite{PDG} of
$g_{\pi\gamma\gamma} \, = \, 0.273$ GeV$^{-1}$  and
$\Gamma_{\pi\gamma\gamma} \, = \, 7.7 \pm \, 0.5 \, \pm \, 0.5$ eV. 

Now we turn to the discussion of the $Q^2$ dependence of 
$F_{\sigma\gamma^\ast\gamma^\ast}$. First we consider the limiting case 
$\omega = 1$, where one of the photons is virtual and the other one is real. 
The corresponding form factor $F_{\sigma\gamma\gamma^\ast}(Q^2)$ is 
plotted in Fig.10. In the numerical calculation all three types of 
diagrams ($\triangle$, bubble and tadpole) are included.
As was already stated before, the
$\triangle$ diagram gives the dominant contribution, whereas
bubble and tadpole diagrams are significantly suppressed.
Using Eq.~(\ref{rad_H}) 
we determine the slope of the $F_{\sigma\gamma\gamma^\ast}$ form factor as
\begin{equation}  
<r^2_{\sigma\gamma}> = 0.40 \,\, {\rm fm}^2 \, .
\end{equation}
For completeness
we also present the results for $F_{\sigma\gamma^\ast\gamma^\ast}$ for
different values of the asymmetry parameter $\omega = 1, \, 3/4, \, 1/2$ 
and $0$ in Fig. 11. 

The coupling constant $g_{\sigma\gamma\gamma}$ and, according to 
Eq. (\ref{dec_H}), the decay width $\Gamma_{\sigma\gamma\gamma}$ 
are given in our approach as:
\begin{eqnarray}\label{sigma_decay}
g_{\sigma\gamma\gamma} \, = \, 0.330 \,\,\, {\rm GeV}^{-1} 
\hspace*{1cm} {\rm and} \hspace*{1cm} 
\Gamma_{\sigma\gamma\gamma} \, = \, 0.26 \,\,\, {\rm KeV} \,. 
\end{eqnarray} 
Again, the dominant contribution to $g_{\sigma\gamma\gamma}$ arises from 
the triangle diagram. The separate contributions of 
the bubble (bub) and tadpole (tad) diagrams to the coupling constant are 
\begin{eqnarray} 
g_{\sigma\gamma\gamma}\bigg|_{\rm bub}  \, = \,  - \, 
0.4 \times 10^{-4} \,\,\, {\rm GeV}^{-1}\,\,\,\, {\rm and}  
\,\,\,\, g_{\sigma\gamma\gamma}\bigg|_{\rm tad}  \, = \,  - \, 
0.2 \times 10^{-4} \,\,\, {\rm GeV}^{-1}\,. 
\end{eqnarray}
A variation of the $\sigma$ meson mass in the region of 
$0 \, \leq \, M_\sigma \, < \, 2 m \, = \, 0.470$ MeV does not have
much influence on the value for $g_{\sigma\gamma\gamma}$ with 
$g_{\sigma\gamma\gamma} \, \simeq \, 0.31 \pm 0.02$ GeV$^{-1}$.   
In the soft meson mass limit $M_\pi^2 = M_\sigma^2 = 0$ we 
approximately reproduce the low-energy theorem: 
\begin{eqnarray}
g_{\pi\gamma\gamma} \, = \, \frac{9}{10} \,\, g_{\sigma\gamma\gamma} 
\, \simeq \, \frac{1}{4 \, \pi^2 \, F_\pi} \,. 
\end{eqnarray} 
The constants $g_{H\gamma\gamma}$ calculated for zero meson mass values, 
$g_{\pi\gamma\gamma} \, = \, 0.263$ GeV$^{-1}$  and 
$g_{\sigma\gamma\gamma} \, = \,  0.292$ GeV$^{-1}$, are rather close  
to the quantities predicted at $M_\pi = 134.98$ MeV and 
$M_\sigma = 385.4$ MeV. 

Taking the recent result for the two-photon decay width of $\sigma $ 
or $f_0(400-1200)$ of $\Gamma(f_0(400-1200) \to \gamma\gamma) = 
3.8 \pm 1.5$ KeV~\cite{PDG,Boglione} at face value, our direct result of 
Eq. (\ref{sigma_decay}) is off by about an order of magnitude. However, 
predictions for the decay width depend rather sensitively, that is to the 
third power, on the value of the scalar meson mass as evident from 
Eq. (\ref{dec_H}). Using our canonical value of 
$g_{\sigma\gamma\gamma} = 0.330$ GeV$^{-1}$, which, as discussed, is fairly
independent of the mass value, and varying the $\sigma$ mass we obtain:  
\begin{equation}
\begin{array}{l|l|l|l|l|l|l|l|l|l|}
\hspace*{1.4cm} M_{\sigma}\,, \,\,\,\,\,\,
{\rm GeV} \,\,\,& \,\,0.4 \,\,& \,\,0.5\,\, & \,\,0.6\,\, & \,\,0.7 \,\,
& \,\,0.8\,\, & \,\,0.9\,\, & \,\,1\,\, & \,\,1.1\,\, & \,\,1.2\,\, \\ 
\hline 
\Gamma(\sigma \to \gamma\gamma)\,, \,\,\,\,\,\, {\rm KeV} \,\,\,& \,\,0.3\,\, 
& \,\,0.6\,\,& \,\, 1 \,\, & \,\,1.6\,\, & \,\,2.3\,\, & \,\,3.3\,\, 
& \,\,4.6\,\, & \,\,6.1\,\, & \,\,7.9 \,\, \\
\end{array}
\end{equation}
The range of prediction can be summarized as
$\Gamma(\sigma \to \gamma\gamma) = 4.1 \pm 3.8$ KeV,
which now is in qualitative agreement with the experimental result.

\subsection{Strong decay $\sigma \to \pi\pi$} 

Based on the definition given
in Sec. \ref{strongdecay}, our numerical result for the coupling constant
of the strong decay $\sigma \to \pi\pi$ is:
\begin{equation}
g_{\sigma\pi\pi} = 1.8 \, {\rm GeV} \, .
\end{equation}
The value we obtain is close     
to the one extracted from the BES experiment \cite{Huo} with  
$g_{\sigma\pi\pi} = 2.0^{+ 0.30}_{- 0.19}$ GeV and to the prediction 
of the linear $\sigma$ model: 
\begin{eqnarray}\label{gspp}
g_{\sigma\pi\pi}=\frac{M_\sigma^2}{F_\pi}=1.6 \,\,\, {\rm GeV}\,.
\end{eqnarray}
In Eq.~(\ref{gspp}) we used the values $M_\sigma = 385.4$ MeV and 
$F_\pi = 92.7$ MeV as determined in our approach.
Using Eq.~(\ref{dec_sigma}) for the decay width of the
$\sigma$ meson we get $\Gamma_\sigma = 173$ MeV.
This value is smaller than the ones 
reported by the E791 ($324_{-40}^{+42}\pm 21$ MeV) and the 
BES ($282_{-50}^{+77}$ MeV) Collaborations. Again, a variation 
of the $\sigma$ meson mass in the region 
$0 \, \leq \, M_\sigma \, < \, 2 m$ leads to a range of predictions for 
$g_{\sigma\pi\pi}(M_\sigma^2) \, = \, 1.96 \, \pm \, 0.73$. The central value 
of $g_{\sigma\pi\pi}(M_\sigma^2)$ corresponds to $M_\sigma \sim 350$ MeV, 
the upper limit to $M_\sigma = 0$, whereas the minimal value is obtained for 
a mass value near $M_\sigma \sim 2 m$. Substituting the results for 
$g_{\sigma\pi\pi}(M_\sigma^2)$ as a function $M_\sigma$ into 
Eq.(\ref{dec_sigma}) we estimate the variation of the $\sigma$ meson width 
as $\Gamma_\sigma = 0 \div 206 $ MeV. The lower limit corresponds obviously 
to the threshold $M_\sigma = 2 M_\pi$ and the largest value to 
$M_\sigma \approx 330$ MeV. 

\subsection{$\sigma$ meson in weak decays} 

Finally we discuss the results for the nonleptonic decays 
$D \to \sigma\pi$ and $B \to \sigma\pi$. We first give our predictions 
for the weak form factors $F_0^{D\sigma}(M_\pi^2)$ and 
$F_0^{B\sigma}(M_\pi^2)$ of Eq. (\ref{weakform}) evaluated at the physical 
point $q^2 = M_\pi^2$:  
\begin{eqnarray}
F_0^{D\sigma}(M_\pi^2) = 0.298  \hspace*{1cm} 
{\rm and} \hspace*{1cm} F_0^{B\sigma}(M_\pi^2) = 0.141 \,. 
\end{eqnarray}
With our results for $F_0^{D\sigma}(M_\pi^2)$ and $F_0^{B\sigma}(M_\pi^2)$ we 
get for the effective weak coupling constants (Eq. (\ref{eq_gds})) 
\begin{eqnarray}
g_{D\sigma\pi} \, = \, 298 \,\,\, {\rm eV} \,\,\,\,\,\, 
{\rm and} \,\,\,\,\,\, 
g_{B\sigma\pi} \, = \, 15.8 \,\,\, {\rm eV} 
\end{eqnarray} 
and finally for the decay widths, originally defined in 
Eq. (\ref{weakwidth}), 
\begin{eqnarray}
\Gamma(D^+\rightarrow\sigma\pi^+)\,=\, 0.90 \times 10^{-12}\,\,\, {\rm MeV} 
\,\,\,\,\,\, {\rm and} \,\,\,\,\,\,  
\Gamma(B^+\rightarrow\sigma\pi^+) = 0.94 \times 10^{-16}\,\,\, {\rm MeV} \,. 
\end{eqnarray}
Our prediction for the decay width $\Gamma(D\rightarrow\sigma\pi)$ 
is in agreement with the lower value of the E791 result~\cite{PDG,E791}
of:  
\begin{eqnarray}
\Gamma(D^+\rightarrow\sigma\pi^+) \, = \, 
(1.32 \pm 0.31) \times 10^{-12} \,\,\, {\rm MeV}\,. 
\end{eqnarray}
We also compare our results to previous theoretical calculations done 
in Refs.~\cite{Dib,Gatto,Deandrea}. A value for $F_0^{D\sigma}(M_\pi^2)$ 
was estimated in Ref.~\cite{Dib} using the $D\to\sigma\pi\to 3\pi$ 
data~\cite{E791} without properly taking into account the rescattering 
effects. The result is $F_0^{D\sigma}(M_\pi^2) \, = \, 0.79 \pm 0.15 $,
which is twice as large as our prediction and also results in a value for
the width $\Gamma(D^+\rightarrow\sigma\pi^+)$ larger than the measured one.
Obviously, the value for $F_0^{D\sigma}(M_\pi^2)$ should be directly 
extracted from the two-body transition $D\to\sigma\pi$. 

In Ref.~\cite{Gatto} the form factor $F_0^{D\sigma}(M_\pi^2)$ was calculated 
(including direct and polar contributions) in a quark model similar 
to ours. The contribution of the direct diagram $0.30 \pm 0.02$  
is close to our result but the contribution of the polar diagram 
is very large $0.22^{+ 0.07}_{-0.01}$. 
With the final result of
$\Gamma(D^+\rightarrow\sigma\pi^+) = 2.3 \times 10^{-12}$ MeV this approach
overestimates the experimental result.
This work was extended in Ref.~\cite{Deandrea} to determine the form factor 
$F_0^{B\sigma}(M_\pi^2)$. The result they obtain is
$F_0^{B\sigma}(M_\pi^2) \, \simeq \,  0.45 \pm 0.15$ 
which, again, is about three times larger than our prediction.

\section{Summary}

In summary, we have applied the relativistic constituent quark model to  
investigate electromagnetic $\pi$ and $\sigma$ meson form factors and  
a variety of strong and weak decay characteristics involving the $\sigma$. 
We start with an effective quark-meson interaction Lagrangian, which is based 
on a linear realization of chiral symmetry. Then we write down the matrix 
elements describing the meson interactions in terms of a set of quark 
diagrams. All model parameters were previously 
determined~\cite{Model}-\cite{EPJdirect} in a fit 
to experimental observables.

In a first step we present a detailed analysis of the electromagnetic form 
factors of the $\pi$ and $\sigma$ mesons. To solidify our model 
considerations, we study the electromagnetic form factor of the charged pion, 
which was recently measured by the Jefferson Lab $F_\pi$  
Collaboration~\cite{Volmer}. We also calculate the form factors which govern  
the transitions $H \to \gamma\gamma$ with $H = \pi^0$ and $\sigma$ 
including different kinematics of the photons. Our results for the pion are  
in good agreement with the recent experimental data~\cite{Gronberg,Volmer}. 
We furthermore give results (analytical formulas and numerics) 
for the asymptotics of the $\pi\to\gamma\gamma$ and $\sigma\to\gamma\gamma$ 
form factors at large values of space-like photon virtualities.
The behavior of the $\pi^0\gamma\gamma^\ast$ form factor nicely coincides
with the prediction of perturbative QCD~\cite{Brodsky_Lepage}.
Expressions for the leading order (LO) $\sim 1/Q^2$
and the next-to-leading (NLO) order $\sim 1/Q^4$ expansion coefficients
in the form factors are calculated at arbitrary values of large photon
virtualities. We complete the analysis by indicating
results for the charge radii.

Based on the chiral symmetry construction we determine the $\sigma$ meson 
mass $M_\sigma = 385.4$ MeV and the total width $\Gamma_\sigma =173$ MeV. 
Both values are lower than the experimental values of the E791~\cite{E791} 
Collaboration, but close to the BES~\cite{BES} results. The predicted
strong coupling constant $g_{\sigma\pi\pi} = 1.8$ GeV is fairly stable
with respect to variations in the $\sigma$ mass, but the total width, due 
to phase space, depends sensitively on this value. Turning to the 
experimental evidence for the $\sigma$ meson we give predictions for the 
decay characteristics in the nonleptonic $D\to\sigma\pi$ and $B\to\sigma\pi$ 
transitions. Both form factors and decays widths are determined. Our result 
for $\Gamma(D^+\rightarrow\sigma\pi^+) = 0.90 \times 10^{-12}$ MeV is 
in agreement with the lower value of data~\cite{PDG,E791}. 
Our prediction for the decay width $\Gamma(B^+\rightarrow\sigma\pi^+)$ is 
$0.94 \times 10^{-16}$~MeV, which is expected to be measured 
in forthcoming experiments.

\newpage 

{\large\bf Acknowledgments}

\vspace*{.5cm}

\noindent 
A.F., Th.G. and V.E.L. thank the DFG grants FA67/25-1 and GRK683 for support.
M.A.I. appreciates the partial support by the DFG grants GRK683 and 
436 RUS17/47/02, the Russian Fund of Basic Research grant No. 01-02-17200 
and the Heisenberg-Landau Program. P.W. thanks the Alexander von Humboldt 
Foundation for financial support.

\newpage 

\centerline{\bf APPENDIX}

\subsection{Methods of calculation for matrix elements} 

We demonstrate the evaluation of matrix elements 
for the example of the derivative of the 
meson mass operator (\ref{Mass-operator}): 
\begin{eqnarray}\label{Pi_tech}
\tilde\Pi^\prime_H(p^2) \, = \, - \, \frac{p^\alpha}{2p^2}\, 
\frac{d}{dp^\alpha} \, \int\!\! \frac{d^4k}{4\pi^2i} \, 
\tilde\Phi^2_H(-k^2) \, {\rm tr} \biggl[\Gamma_H S_1(\not\! k+w_{21} \not\!p) 
\Gamma_H S_2(\not\! k-w_{12} \not\!p) \biggr] \, .  
\end{eqnarray}  
The technique we use is based on the three main ingredients:  
\begin{itemize}
\item  use of the Laplace transform of the vertex function 
$$
\tilde\Phi(z)=\int\limits_0^\infty\! ds\, \Phi_L(s)\, e^{-sz} \, ,
$$
\vspace{-0.2cm}
\item the  $\alpha$-transform of the denominator
$$
\frac{1}{m^2-(k+p)^2}=\int\limits_0^\infty\! d\alpha \,  
e^{-\alpha (m^2-(k+p)^2)} \, ,
$$
\vspace{-0.2cm}
\item and the differential representation of the numerator
$$
\left(\,m+\not\! k +\not\! p \right)\, e^{kq}=
\left(\,m+\gamma^\mu \frac{\partial}{\partial q^\mu} 
+ \not\! p\,\right)\, e^{kq} \, .
$$
\end{itemize}
After straightforward algebra we get expressions for the 
coupling constants $h_H = 1/\tilde\Pi^\prime_H(m_H^2)$ 
of pseudoscalar $(H=P)$ and 
scalar $(H=S)$ mesons, which are given by 
\begin{eqnarray}
h_H^{-1} &=& \frac{1}{4}\int\limits_0^\infty \frac{dt t}{A^3} 
\int\limits_0^1 d\alpha \biggl\{ - [\tilde\Phi^2(z_0)]^\prime 
\, B \biggl( d_H \frac{m_1 m_2}{\Lambda_H^2} \, + 
\, \frac{m_H^2}{\Lambda_H^2} \, \frac{A (1 - \Delta^2) + B}{4 A^2} 
\biggr)\\
&+& \tilde\Phi^2(z_0) \biggl(1 - \Delta^2 + \frac{3B}{A} \biggr) 
\biggr\} \, , \nonumber
\end{eqnarray}
where $[\tilde\Phi^2(z_0)]^\prime = d[\tilde\Phi^2(z_0)]/dz_0$, 
$d_H = 1$ for pseudoscalar mesons and 
$d_H = - 1$ for scalar mesons. We also define
\begin{eqnarray}
z_0 &=& \frac{B}{4} [(m_1+m_2)^2 - m_H^2] 
+ \frac{m_H^2 t^2}{4A} (2\alpha - 1 + \Delta)^2 \, , \nonumber\\
& &\nonumber\\
A &=& 1+t, \,\,\,\,\,  
B = t [4\alpha(1-\alpha) A +  (2 \alpha - 1 + \Delta)^2], \,\,\,\,\,  
\Delta = w_{21} - w_{12} \, . \nonumber
\end{eqnarray}

In analogy we consider the basic vertex function (\ref{Lambda_Trans}):  
\begin{eqnarray} 
\Lambda^{12; 13; 23}(p_1,p_2) &=&
\frac{3}{4\pi^2} g_{H_{13}} g_{H_{23}} g_{H_{12}} 
I^{12; 13; 23}(p_1,p_2) \\ 
I^{12; 13; 23}(p_1,p_2) &=& - \, \int\!\! \frac{d^4k}{4\pi^2i} \, 
\tilde\Phi_{H_{13}}( - (k+w_{13}\,p_1)^2) \, 
\tilde\Phi_{H_{23}}( - (k+w_{23}\,p_2)^2)\nonumber\\
&\times&\tilde\Phi_{H_{12}}( - (k+w_{12}\,p_1+w_{21}\,p_2)^2) 
{\rm tr} [ S_2(\not\! k+\not\! p_2) \Gamma_{H_{12}}  
S_1(\not\! k+\not\! p_1)\Gamma_{H_{13}} 
S_3(\not\! k)\Gamma_{H_{23}} ]\nonumber
\end{eqnarray} 
Here we indicate the resulting expression for the
Gaussian meson-quark vertex function 
$\tilde\Phi_H(k_E^2) = \exp(-k_E^2/\Lambda_H^2)$: 
\begin{eqnarray} 
I^{12; 13; 23}(p_1,p_2) &=&
\int\limits_0^\infty \frac{dt \, t}{A^2} \int\limits_0^1 d^3\alpha  
\delta\biggl(1 - \sum\limits_{i=1}^3 \alpha_i\biggr)  
\exp\biggl( - \frac{z}{\Lambda_{\rm red}^2}\biggr) 
\biggl( \frac{C_1}{\Lambda_{\rm red}^2} - 
\frac{C_2}{2 A} \biggr) \, ,   
\end{eqnarray}
where 
\begin{eqnarray}
C_1 &=& \frac{1}{4} {\rm tr} [(m_2 + \not\!\! D_2) \Gamma_{12}  
(m_1 + \not\!\! D_1) \Gamma_{13} (m_3 + \not\!\! D_3) 
\Gamma_{23}] \,, \nonumber\\
& &\nonumber\\
C_2 &=& \frac{1}{4} {\rm tr} [ 
\gamma^\sigma \Gamma_{12} \gamma_\sigma  
\Gamma_{13} (m_3 + \not\!\! D_3) \Gamma_{23} +  
\gamma^\sigma \Gamma_{12} (m_1 + \not\!\!D_1)  
\Gamma_{13} \gamma_\sigma \Gamma_{23} + 
(m_2 + \not\!\! D_2) \Gamma_{12} \gamma^\sigma  
\Gamma_{13} \gamma_\sigma \Gamma_{23} ] ,\nonumber \\
& &\nonumber\\
   D_1 &=& p_1 - p_t, \,\,\,\,\, 
   D_2 = p_2 - p_t, \,\,\,\,\, 
   D_3 = - p_t, \,\,\,\,\, p_t = r_1 p_1 + r_2 p_2 \, ,\nonumber\\
r_1 &=& \frac{1}{A} ( t\alpha_1 + w_{13} s_{13} + w_{12} s_{12} ), 
\,\,\,\,\,  
r_2 = \frac{1}{A} ( t\alpha_2 + w_{23} s_{23} + w_{12} s_{21} ) \,, 
\nonumber \\
& &\nonumber\\
z &=& t [ \sum\limits_{i=1}^{3} \alpha_i m_i^2 - \alpha_1 p_1^2 
- \alpha_2 p_2^2 ] + A [ (r_1 + r_2) (r_1 p_1^2 + r_2 p_2^2) - 
r_1 r_2 p_3^2 ]  \nonumber\\
& &\nonumber\\
  &-& p_1^2 (w_{13}^2 s_{13} + w_{12} s_{12}) 
- p_2^2 (w_{23}^2 s_{23} + w_{21} s_{12}) 
+ p_3^2 w_{12} w_{21} s_{12} \, , \nonumber\\ 
& &\nonumber\\ 
\frac{1}{\Lambda_{\rm red}^2} &=& 
\frac{1}{\Lambda_{H_{12}}^2} +   \frac{1}{\Lambda_{H_{13}}^2} +   
\frac{1}{\Lambda_{H_{23}}^2}, \,\,\,\,\, 
s_{ij} = \frac{\Lambda_{red}^2}{\Lambda_{ij}^2} \, . \nonumber
\end{eqnarray}
All further calculations are done by using computer programs written
in FORM for the manipulations of Dirac matrices and in FORTRAN
for the final numerical evaluations.

\subsection{Feynman rules for nonlocal electromagnetic vertices}  

In the following we derive the Feynman rules for the nonlocal vertices
of Figs.3a and 3b.  
These vertices contain the path integral over the gauge field 
$I(x,y,P)$. The crucial point is to calculate the expression 
\begin{eqnarray}
\tilde\Phi(\partial_x^2) e^{ipx} [I(x,y,P)]^N
\end{eqnarray}
using Eq.(\ref{path2}). The case $N=1$ corresponds to the vertex of Fig.3a 
and $N=2$ to the vertex of Fig.3b. 

First, we consider the case $N=1$. It is readily seen that 
\begin{eqnarray}
\tilde\Phi(\partial_x^2) e^{ipx} I(x,y,P) = 
e^{ipx} \tilde\Phi({\cal D}_x^2) I(x,y,P)
\end{eqnarray}
where ${\cal D}_x \equiv \partial_x + i p$. Thus we have to calculate
\begin{eqnarray}
\tilde\Phi({\cal D}_x^2) I(x,y,P) \, = \, \sum\limits_{n=0}^{\infty} \, 
\frac{\tilde\Phi^{(n)}(0)}{n!} \, {\cal D}_x^{2n} \, I(x,y,P) \, . 
\end{eqnarray}
One finds that 
\begin{eqnarray}
{\cal D}_x^2 I(x,y,P) \, = \, [\partial_x A(x) + 2i p A(x)] - p^2 I(x,y,P) 
\equiv L(A) - p^2 I(x,y,P) 
\end{eqnarray}
where $L(A)\equiv \partial_x A(x) + 2i p A(x)$. 

Iteration of the last expression
\begin{eqnarray}
({\cal D}_x^2)^2 I(x,y,P) &=& ({\cal D}_x^2 - p^2) L(A) + (-p^2)^2 I(x,y,P)\\
({\cal D}_x^2)^3 I(x,y,P) &=& ({\cal D}_x^4 - {\cal D}_x^2 p^2 + p^4) L(A) 
+ (-p^2)^3 I(x,y,P) \nonumber\\ 
&\ldots&\nonumber\\
({\cal D}_x^2)^n I(x,y,P) &=& \sum\limits_{k=0}^{n-1} \, 
({\cal D}_x^2)^{n-1-k} \, (- p^2)^k \, L(A) \, + \, (-p^2)^n I(x,y,P)
\nonumber\\
 &=& n \, \int\limits_0^1 \, dt \, 
[{\cal D}_x^2 t \, - \, p^2 (1-t) ]^{n-1} \, L(A) \, + \, (-p^2)^n I(x,y,P)
\nonumber
\end{eqnarray}
finally leads to 
\begin{eqnarray}\label{EqN1_last}
\tilde\Phi({\cal D}_x^2) I(x,y,P) &=& \int\limits_0^1 \, dt \, 
\tilde\Phi^\prime [{\cal D}_x^2 t \, -  \, p^2 (1-t) ] \, L(A) \, + 
\, \tilde\Phi(-p^2) \, I(x,y,P)\\
&=&\int \frac{d^4 q}{(2\pi)^4} \,  A_\mu(q) \, 
\bigg\{ i (2p+q)^\mu e^{i q x } \int\limits_0^1 dt 
\tilde\Phi^\prime[ - (p+q)^2 t \, - \, p^2 (1-t) ] \nonumber\\
&+& \tilde\Phi(- p^2) \int\limits_y^x dz^\mu e^{iqz} \biggr\}
\nonumber \,  
\end{eqnarray}
where $A_\mu(q)$ is the Fourier-transform of the electromagnetic field and 
$\tilde\Phi^\prime(z) = d\tilde\Phi(z)/dz$. 
The last term in Eq.(\ref{EqN1_last}) containing an integration 
from $y$ to $x$ (path integral) vanishes due to the delta function 
$\delta(x-y)$ in the Lagrangian. 

In complete analogy, we also obtain for the nonlocal vertex for the case $N=2$: 
\begin{eqnarray}\label{EqN2_last}
\tilde\Phi({\cal D}_x^2) I^2(x,y,P) &=& 
\int \frac{d^4 q_1}{(2\pi)^4} \, \int \frac{d^4 q_2}{(2\pi)^4} 
\,  A_\mu(q_1) \, A_\nu(q_1) \\
&\times& \bigg\{ 2 g^{\mu\nu} \, e^{i (q_1+q_2) x} 
\int\limits_0^1 dt 
\tilde\Phi^\prime[ - (p+q_1+q_2)^2 t \, - \, p^2 (1-t) ] \nonumber\\
&-&(2p + q_1)^\mu (2p + 2q_1 + q_2)^\nu \, e^{i (q_1 + q_2) x} \, 
\int\limits_0^1 dt \, t \, \int\limits_0^1 dl \,  \nonumber\\
&\times&\tilde\Phi^{\prime\prime}[ - l( t(p + q_1 + q_2)^2  \, + \, p^2 (1-t)) 
- (1 - l) ( t(p + q_1)^2 + (1 - t) p^2) ] \nonumber\\
&-&(2p + q_2)^\nu (2p + 2q_2 + q_1)^\mu \, e^{i (q_1+q_2) x} \, 
\int\limits_0^1 dt \, t \, \int\limits_0^1 dl \, \nonumber\\
&\times&\tilde\Phi^{\prime\prime}[ - l( t(p + q_1 + q_2)^2  \, + \, p^2 (1-t)) 
- (1 - l) ( t(p + q_2)^2 + (1 - t) p^2) ] \nonumber\\
&+& i (2p+q_1)^\mu e^{i q_1 x} \int\limits_0^1 dt 
\tilde\Phi^\prime[ - (p+q_1)^2 t \, - \, p^2 (1-t) ] 
\int\limits_y^x dz^\mu e^{iq_2 z} \nonumber\\
&+& i (2p+q_2)^\nu e^{i q_2 x } \int\limits_0^1 dt 
\tilde\Phi^\prime[ - (p+q_2)^2 t \, - \, p^2 (1-t) ] 
\int\limits_y^x dz^\mu e^{iq_1 z} \nonumber\\
&+& g^{\mu\nu} \tilde\Phi(- p^2) \int\limits_y^x dz^\mu e^{iq_1 z}  
               \int\limits_y^x dw^\nu e^{iq_2w} \biggr\} \, ,
\nonumber \,  
\end{eqnarray}
where $\tilde\Phi^{\prime\prime}(z) = d^2\tilde\Phi(z)/dz^2$. 
Again, the last three terms in Eq.(\ref{EqN2_last}) containing integrations 
from $y$ to $x$ vanish due to the delta function.

\subsection{Form factors characterizing the local $\sigma\to\gamma\gamma$ diagram} 

In the following we give the full analytical expressions for the form factors
of Eq. (\ref{coeff_alp1}):

\begin{eqnarray}\label{coeff_res1} 
F^{\sigma}_{\rm L} &=& \frac{m}{\lambda^2}
\left\{
-2\, (p^2-q_1^2-q_2^2)\,\lambda
-2\, L_1\,q_1^2\,(\lambda-6\, q_2^2\, (q_2^2-q_1^2-p^2))
\right.
\\
&-& 2\, L_2\,q_2^2\,(\lambda-6\, q_1^2\, (q_1^2-q_2^2-p^2))
\nonumber\\
&+& 
2\, L_3 (q_1^6+q_2^6-q_1^4\, q_2^2-q_1^2\, q_2^4-2\, q_1^4\, p^2-2\, q_2^4\, p^2
           +q_1^2\, p^4+q_2^2\, p^4+ 8\, q_1^2\, q_2^2\, p^2)
\nonumber\\
&+& I_{\rm L}(p^2,q_1^2,q_2^2) \,(p^2-q_1^2-q_2^2)\,(4\, m^2\, \lambda-p^6
    +q_1^6+q_2^6-q_1^4\, q_2^2-q_1^2\, q_2^4  
\nonumber\\\,
&& 
\left.
-3\, q_1^4\, p^2-3\, q_2^4\, p^2+3\, q_1^2\, p^4+3\, q_2^2\, p^4
+10\, q_1^2\, q_2^2\, p^2)
\right\} \, ,
\nonumber\\
&&\nonumber\\
G^{\sigma}_{\rm L} &=& \frac{m}{\lambda^2}
\left\{
4 \,\lambda+4\, L_1 (\lambda-3\, q_1^2\, (q_1^2+q_2^2+p^2))
         +4\, L_2\, (\lambda-3\, q_2^2\, (q_1^2+q_2^2+p^2))
\right.
\label{coeff_res2}\\
&&       +4\, L_3\, (\lambda-3\, p^2\, (q_1^2+q_2^2+p^2))
\nonumber\\
&-& 2\, I_{\rm L}(p^2,q_1^2,q_2^2) \, (4\, m^2\, \lambda+p^6
    +q_1^6+q_2^6-q_1^4\, q_2^2-q_1^2\, q_2^4-q_1^4\, p^2-q_2^4\, p^2
\nonumber\\
&& 
\left. - q_1^2\, p^4-q_2^2\, p^4+6\, q_1^2\, q_2^2\, p^2)
\right\}
\nonumber
\end{eqnarray}
where $I_{\rm L}(p^2,q_1^2,q_2^2)$ is the integral 
defined in Eq. (\ref{I_L}). Here we use

\begin{equation}
L_i=\int\limits_0^1\! dx \ln\left[1-x(1-x)\,\frac{q^2_i}{m^2}\right]
= -2+2\,\sqrt{\frac{4\,m^2-q^2_i}{q^2_i}}\, 
{\rm ArcTan}\left(\sqrt{\frac{q^2_i}{4\,m^2-q^2_i}}\right)
\end{equation}
with $q_3\equiv p$.

\subsection{Two-loop integral representation and power expansion of $I_{\rm NL}$}

The integral $I_{\rm NL}$ of Eq. (\ref{pion-nonloc}), which is related to
the $\pi^0 \gamma \gamma $ form factor, can be reduced to the two-loop integral 
\begin{eqnarray}\label{pi_NL1}
I_{\rm NL} \, = \,
-\,\int\limits_0^\infty\! dt \frac{t}{1+t}\int\limits_0^1\! dx
\left\{
 \frac{\tilde\Phi(z_p)}{t\,A_1(x)+\Delta_1} \, 
-\frac{(1-\alpha_0) \, \tilde\Phi(z_{q_2})}{t\,A_2(x)+\Delta_1} \, 
-\frac{\alpha_0 \, \tilde\Phi(z_{q_1})}{t\,A_3(x)+\Delta_1} \, 
\right\}\,, 
\end{eqnarray}
where

$$
\def\arraystretch{1.5}
\begin{array}{ll} 
z_p=\frac{t^2}{1+t}\,D_p+\frac{t}{1+t}\,\Delta(1) &
D_p=m^2-x\,(1-x)\,p^2 
\\
z_{q_i}=\frac{t^2}{1+t}\,D_{q_i}+\frac{t}{1+t}\,\Delta(x) &
D_{q_i}=m^2-x\,(1-x)\,q_i^2 
\\
\Delta(x)=\Delta_0+x\,\Delta_1 &
\Delta_0=m^2+p^2/4-(q_1^2+q_2^2)/2 
\\
A_1(x)=x\cdot\lambda^{1/2}+q_1^2-\alpha_0\,p^2 &
\Delta_1= -p^2/2+(q_1^2+q_2^2)/2 
\\
A_2(x)=(1-\alpha_0)\,x\cdot\lambda^{1/2}-(1-\alpha_0)\,q_1^2
-\alpha_0\,q_2^2 &
\\
A_3(x)=-\alpha_0\,x\cdot\lambda^{1/2}-(1-\alpha_0)\,q_1^2
-\alpha_0\,q_2^2 &
\\
\end{array}
$$

With the constraint that $p^2\le 4\,m^2$ and $q^2_i\le 0$
the variable $t$ is replaced by $u=t^2\,D/(1+t)+t\,\Delta/(1+t)$ in the
integral of Eq.~(\ref{pi_NL1}):

\begin{eqnarray}\label{pi_NL2}
I_{\rm NL} \, = \, \int\limits_0^\infty\! 
du \, \tilde\Phi(u) \int\limits_0^1\! dx \, 
\biggl\{ &-& \, \frac{1}{R_p(1)} \,\,
\frac{u - \Delta(1) + R_p(1)}{[u-\Delta(1) + R_p(1)]
           \cdot A_1(x) + 2 \, D_p \, \Delta_1}\, \\
&+& \,\, \frac{1}{R_{q_2}(x)} \, 
 \frac{(1-\alpha_0) \, [u - \Delta(x) + R_{q_2}(x)]}
      {[u-\Delta(x) + R_{q_2}(x)] \cdot A_2(x) + 2\, D_{q_2} \, \Delta_1} \,
\nonumber\\ 
&+& \,\, \frac{1} {R_{q_1}(x)} \, \frac{\alpha_0 \, [u-\Delta(x) + R_{q_1}(x)]}
{[u - \Delta(x) + R_{q_1}(x)] \cdot A_3(x) + 2 \, D_{q_1} \, \Delta_1}
\biggr\} \, ,\nonumber
\end{eqnarray}
where $R_p(x) = \sqrt{(u-\Delta(x))^2 + 4 \,u \, D_p}$. 

Next we perform a power expansion of $I_{\rm NL}$ in the limit 
$p^2=0$. For this kinematics we have 
\begin{eqnarray}
D &=&
m^2+\frac{1}{2}\alpha_3\,Q^2\,(\alpha_1\,(1+\omega)+\alpha_2\,(1-\omega))\,,
\\
\Delta &=& m^2+\frac{1}{2}\alpha_3\,Q^2\,. \nonumber
\end{eqnarray}

By scaling the variable $\alpha_3\to \alpha_3/Q^2$ one finds 

\begin{eqnarray}\label{asym-nonloc1}
I_{\rm NL} = - \frac{1}{Q^2}  
\int\limits_0^\infty\! dt\,\left(\frac{t}{1+t}\right)^2
\int\!d^3\alpha \, \delta\left(1-\alpha_1-\alpha_2-\frac{\alpha_3}{Q^2}\right)
\tilde\Phi^{\,\prime}\left(t\,m^2+\frac{t \, \alpha_3 \, W}{2\,(1+t)}\right) \, ,
\end{eqnarray}
where
\begin{equation}\label{wdef}
W=1+t\,[\alpha_1\,(1+\omega)+\alpha_2\,(1-\omega)] \, .
\end{equation} 
From the last expression
it is easy to determine the asymptotics in the leading order (LO) with 
\begin{eqnarray}\label{coeff_pi_asym}
\left(I_{\rm NL}\right)_{\,\rm LO} \, = \, 
\,\frac{1}{Q^2} \, R_\pi^{\rm LO}(\omega)\,. 
\end{eqnarray}

By using the formulas

\begin{eqnarray}
&&
\int \! d^2\alpha \, \delta^\prime(1-\alpha_1-\alpha_2) \, F(\alpha_1,\alpha_2) 
= \int \! d^2\alpha \, \delta(1-\alpha_1-\alpha_2) \, 
\left(F(0,1) + F^\prime_{\alpha_1}(\alpha_1,\alpha_2)\right)\,,\nonumber\\
&&\\
&&
\int\limits_0^\infty\! dt \, g(t)\int\limits_{t\,m^2}^\infty\! d\tau \, f(\tau) 
= m^2 \,\int\limits_0^\infty\! dt \, f(t\,m^2)\,
\int\limits_0^t\! d\tau \, g(\tau)\,,\nonumber
\end{eqnarray}
where $F^\prime_{\alpha_1}(\alpha_1,\alpha_2) = 
\partial F(\alpha_1,\alpha_2)/\partial\alpha_1$,   
the next-to-leading (NLO) term is obtained as
\begin{eqnarray}
\left(I_{\rm NL}\right)_{\,\rm NLO} = 
\frac{1}{Q^4} \, R_\pi^{\rm NLO}(\omega)\,. 
\end{eqnarray} 
Note that both integrals $R_\pi^{\rm LO}(\omega)$ and 
$R_\pi^{\rm NLO}(\omega)$ are defined in Eq.~(\ref{R_pi}). 

\subsection{Form factors characterizing the gauge-invariant part of the 
nonlocal triangle $\sigma\to\gamma\gamma$ diagram}

Here we give the full analytical expressions for the form factors
of Eqs. (\ref{coeff_trn_alp1}) and (\ref{coeff_trn_alp2}):

\begin{eqnarray}\label{coeff_trn1}
F^{\sigma}_{\triangle_\perp} &=& \frac{m}{\lambda^2}
\left\{8\,(S_q-S_p)\,(\lambda+6\,q_1^2\,q_2^2)
+2\, L_1^{\rm NL}\,q_1^2\,(\lambda-12\, q_2^4+12\,q_2^2\,p^2)
\right.
\\
&& 
+2\, L_2^{\rm NL}\,q_2^2\,(\lambda-12\, q_1^4+12\,q_1^2\,p^2)
\nonumber\\
&& 
-2\, L_3^{\rm NL}\,(q_1^6+q_2^6-q_1^4\, q_2^2-q_1^2\, q_2^4
            -2\, q_1^4\, p^2-2\, q_2^4\, p^2
            +q_1^2\, p^4+q_2^2\, p^4+ 8\, q_1^2\, q_2^2\, p^2)
\nonumber\\
&&
+6\,V_{11}\,q_1^2\,(q_1^2+q_2^2-p^2)(q_1^2+3\,q_2^2-p^2)
-6\,V_{12}\,q_1^2\,(q_1^2-q_2^2-p^2)(q_1^2+q_2^2-p^2)
\nonumber\\
&& 
+6\,V_{21}\,q_2^2\,(q_1^2+q_2^2-p^2)(3\,q_1^2+q_2^2-p^2)
+6\,V_{22}\,q_2^2\,(q_1^2+q_2^2-p^2)(   q_1^2-q_2^2+p^2)
\nonumber\\
&&
+ I_{\rm NL}(p^2,q_1^2,q_2^2)  \, (p^2-q_1^2-q_2^2)\,(4\, m^2\, 
\lambda - p^6+q_1^6+q_2^6-q_1^4\, q_2^2-q_1^2\, q_2^4  
\nonumber\\
&& 
\left.
-3\, q_1^4\, p^2-3\, q_2^4\, p^2+3\, q_1^2\, p^4+3\, q_2^2\, p^4
+10\, q_1^2\, q_2^2\, p^2)
\right\}
\nonumber\\
&&\nonumber\\
G^{\sigma}_{\triangle_\perp} &=& \frac{m}{\lambda^2}
\left\{
\frac{2}{q_1^2\,q_2^2}\,\,(S_q-S_p)\,\,(q_1^2+q_2^2-p^2)
                                       (\lambda+12\,q_1^2\,q_2^2)
\right.
\label{coeff_trn2}\\
&&
+4\, L^{\rm NL}_1 (q_1^4-2\,q_2^4-2\,p^4-5\,q_1^2\, q_2^2
                      +q_1^2\,p^2+4\,q_2^2\,p^2)
\nonumber\\
&& 
-4\, L^{\rm NL}_2 (2\,q_1^4-q_2^4+2\,p^4+5\,q_1^2\, q_2^2
                      -4\,q_1^2\,p^2-\,q_2^2\,p^2)
\nonumber\\
&& 
-4\, L^{\rm NL}_3 (q_1^4+q_2^4-2\,p^4-2\,q_1^2\,q_2^2
                      +q_1^2\,p^2+\,q_2^2\,p^2)
\nonumber\\
&& 
+2\,V_{11}\frac{1}{q_2^2}(q_1^2+3\,q_2^2-p^2)(\lambda+6\,q_1^2\,q_2^2)
-2\,V_{12}\frac{1}{q_2^2}(q_1^2-q_2^2-p^2)(\lambda+6\,q_1^2\,q_2^2)
\nonumber\\
&& 
+2\,V_{21}\frac{1}{q_1^2}(3\,q_1^2+q_2^2-p^2)(\lambda+6\,q_1^2\,q_2^2)
+2\,V_{22}\frac{1}{q_1^2}(q_1^2-q_2^2+p^2)(\lambda+6\,q_1^2\,q_2^2)
\nonumber\\
&&
-2\, I_{\rm NL}(p^2,q_1^2,q_2^2) \, (4\, m^2\, \lambda+p^6
    +q_1^6+q_2^6-q_1^4\, q_2^2-q_1^2\, q_2^4-q_1^4\, p^2-q_2^4\, p^2
\nonumber\\
&& 
\left.
-q_1^2\, p^4-q_2^2\, p^4+6\, q_1^2\, q_2^2\, p^2)
\right\}
\nonumber
\end{eqnarray}
where $I_{\rm NL}(p^2,q_1^2,q_2^2)$ is the three-dimensional integral 
of Eq. (\ref{pion-nonloc}). Above we introduce the terms

\begin{eqnarray}\label{trn-int}
S_r &=& \int\limits_0^\infty\! dt \frac{t}{1+t}
\left(m^2-\frac{1}{(1+t)^2} \frac{r^2}{4}\right) \, \tilde\Phi(z_r)\,,
\\
&&\nonumber\\
L_i^{\rm NL}&=&
\int\limits_0^\infty\!dt \frac{t}{(1+t)^2}
\int\limits_0^1\! dx\,\tilde\Phi(z_i)\,,
\\
&&\nonumber\\
V_{i1}&=&\int\limits_0^\infty\!dt \frac{t^2}{(1+t)^3}
\int\limits_0^1\! dx\,x \, \tilde\Phi(z_i)\,, \\
&&\nonumber\\
V_{i2}&=&
\int\limits_0^\infty\!dt \frac{t^2}{(1+t)^3}
\int\limits_0^1\! dx\,(1-x)\,\tilde\Phi(z_i)\,,
\end{eqnarray}
where

\begin{eqnarray*}
z_r &=& m^2\,t-\frac{r^2}{4}\frac{t}{1+t}\,, \hspace{1cm} (r=p,q)\,,
\\
z_i &=& \frac{t^2}{1+t}\left(m^2-x\,(1-x)\,q_i^2\right)
+\frac{t}{1+t}
\left(m^2-\frac{p^2}{4}+\frac{x}{2}(p^2-q_1^2-q_2^2)\right)\,,
\hspace{1cm} (i=1,2)\,,
\\
z_3 &=& \frac{t^2}{1+t}(m^2-x\,(1-x)\,p^2)
+\frac{t}{1+t}\left(m^2-\frac{p^2}{4}\right)\,.
\end{eqnarray*}

The asymptotics of these expressions can be derived
from their representation in a manner as discussed for the 
$\pi\gamma\gamma$ form factor in Appendix D. Here we obtain

\begin{eqnarray}\label{coeff_asym}
\left(F^{\sigma}_{\triangle_\perp}\right)_{\,\rm LO} &=&
- \, \frac{m}{Q^2 \, \omega^2} \, \biggl\{ R_\pi^{\rm LO}(\omega) 
\, - \, \frac{2}{3} \, R_\pi^{\rm NLO}(\omega) \biggr\}\,, \\ 
\left(F^{\sigma}_{\triangle_\perp}\right)_{\,\rm NLO} &=& - 
\frac{m^3}{Q^4} \, R_{\sigma 1}^{\rm NLO}(\omega)\,, \nonumber\\
\left(G^{\sigma}_{\triangle_\perp}\right)_{\,\rm LO} &=&
-\frac{2}{Q^2} 
\left(F^{\sigma}_{\triangle_\perp}\right)_{\,\rm LO}\,, \nonumber\\
\left(G^{\sigma}_{\triangle_\perp}\right)_{\,\rm NLO} &=& 
- \frac{2}{Q^2} \, 
\left(F^{\sigma}_{\triangle_\perp}\right)_{\,\rm NLO} \, - \, 
\frac{8 \, m}{Q^6} \, R_{\sigma 2}^{\rm NLO}(\omega)\,. \nonumber
\end{eqnarray}
The integrals $R_{\sigma 1}^{\rm NLO}(\omega)$ and 
$R_{\sigma 2}^{\rm NLO}(\omega)$ are defined in Eq.~(\ref{R_sigma}). 
The last simple relation between 
$\left(F^{\sigma}_{\triangle_\perp}\right)_{\,\rm LO}$ and 
$\left(G^{\sigma}_{\triangle_\perp}\right)_{\,\rm LO}$ is 
derived with the help of the identity 
\begin{eqnarray}
t \, \omega \, ( \alpha_1 \, - \, \alpha_2 ) \, = \, W \, - \, 
( 1 \, + \, t \, (\alpha_1 \, + \, \alpha_2 ) \, ) \, ,
\end{eqnarray}
where W is defined in Eq. (\ref{wdef}), 
and the exchange symmetry $\alpha_1 \leftrightarrow \alpha_2$. 

\subsection{Form factors characterizing the gauge-invariant part of 
the nonlocal bubble and tadpole $\sigma\gamma\gamma$ diagrams}

The gauge invariant parts of the nonlocal $\sigma\gamma\gamma $ bubble (bub)
diagram introduced in Eq. (\ref{sigma-bub}) are written as:
\begin{eqnarray}
F^{\sigma}_{\rm bub_\perp} &=&
\frac{m}{2}\int\limits_0^\infty\!dt \frac{t^2}{(1+t)^4}
\int\limits_0^1\! d\tau\,\tau\int\limits_0^1\! dx\,
(1-2\,x)(\tilde\Phi^{\,\prime}(z_1)+\tilde\Phi^{\,\prime}(z_2))\,,
\label{coeff_bub1}\\
&&\nonumber\\
G^{\sigma}_{\rm bub_\perp} &=& - \frac{m}{q_1^2\,q_2^2}
\int\limits_0^\infty\!dt \frac{t}{(1+t)^3}
\int\limits_0^1\! d\tau\,\int\limits_0^1\! dx\,
\left\{
\tilde\Phi(z_1) + \tilde\Phi(z_2)
\right.
\label{coeff_bub2}\\
&+& \left. \frac{1}{4}\,(p^2-q_1^2-q_2^2)\frac{t\,\tau}{1+t}(1-2\,x)
(\tilde\Phi^\prime(z_1) + \tilde\Phi^\prime(z_2))
\right\}\,,\nonumber
\end{eqnarray}
where

\begin{eqnarray}
z(q_1,q_2) &=& t\,\left(m^2-x\,(1-x)\,q_1^2\right)
+\frac{t \, \tau}{1+t} \, (1-2\,x) \, \frac{p^2 - q_1^2 - q_2^2}{4}
\nonumber\\
&-& \frac{\tau \, (1 - \tau + t)}{1 + t} \, \frac{q_2^2}{4}
- \frac{t}{1 + t} \, (1 - 2\,x)^2 \, \frac{q_1^2}{4} \, ,\nonumber\\
z_1 &=& z(q_1,q_2), \,\,\,\,\, z_2 \, = \, z(q_2,q_1) \, .\nonumber
\end{eqnarray}

The final expressions for the tadpole (tad) diagram,
originally defined in Eq. (\ref{sigma-tad}), are

\begin{eqnarray}
F^{\sigma}_{\rm tad_\perp} &=&
- \frac{m}{4}\int\limits_0^\infty\!dt \frac{1}{(1+t)^3}
\int\limits_0^1\! d\tau\,\tau^2\,\int\limits_0^1\! dl\,l
\label{coeff_tad1}\\
&\times&
\bigg\{
\tilde\Phi^\prime(z_1) + \tilde\Phi^\prime(z_2) -
\tilde\Phi^\prime(z_3) - \tilde\Phi^\prime(z_4)\bigg\}
\left(1-\frac{\tau}{1+t}\right)\,,
\nonumber\\
&&\nonumber\\
G^{\sigma}_{\rm tad_\perp} &=&
\,\frac{m}{8\,q_1^2\,q_2^2}\int\limits_0^\infty\!dt 
\frac{1}{(1+t)^2}
\int\limits_0^1\! d\tau\,\int\limits_0^1\! dl\,
\bigg\{4\,(\tilde\Phi(z_p) + \tilde\Phi(z_q))\label{coeff_tad2}\\
&+&
\frac{l\,\tau^2\,(1+t-\tau)}{(1+t)^2}
(q_1^2+q_2^2-p^2)\,
(\tilde\Phi^\prime(z_1) + \tilde\Phi^\prime(z_2) -
\tilde\Phi^\prime(z_3) - \tilde\Phi^\prime(z_4))\nonumber\\
&-&
\frac{4}{1+t}\,(\tilde\Phi(z_1) + \tilde\Phi(z_2) -
\tilde\Phi(z_3) + \Phi^\prime(z_4))\bigg\}\,,
\nonumber
\end{eqnarray}
where

\begin{eqnarray*}
z(p) &=& m^2\,t \, - \, \frac{p^2}{4}\,\tau\,\frac{ 1 + t - \tau}{1 + t}\,, \,\,\,\,\, 
z_p \, = \, z(p)\,, \,\,\, z_q \, = \, z(q)\,, \,\,\,\,\,   
z_i  = m^2\,t - W_i - R_i \frac{ 1 + t - \tau}{1 + t}\,, \\
W(q) &=& \frac{q^2}{4}\, \tau\, l \, (1\, - \, l)\,, \,\,\,\,\, 
R(q_1,q_2,l) \, = \, \frac{\tau^2}{4}\,\left(l\,[p^2 - q_1^2 - q_2^2] \, + \, l^2 \, 
q_1^2 \, + \, q_2^2\right)\,,\\
W_1 & = & W_3 \, = \, W(q_1)\,, \,\,\,\,\,  
W_2 \, = \, W_4 \, = \, W(q_2)\,, \\ 
R_1 &=& R(q_1,q_2,l)\,, \,\,\,\,\, 
   R_2 \, = \, R(q_2,q_1,l)\,, \,\,\,\,\, 
   R_3 \, = \, R(q_1,q_2,-l)\,, \,\,\,\,\, 
   R_4 \, = \, R(q_2,q_1,-l)\,.
\end{eqnarray*}

\vfill 

\newpage

\begin{table}[ht]
\caption{
Leptonic decay constants $f_P$ (MeV) used in the least-square fit.
Data are either taken from PDG  $\protect\cite{PDG}$ or
from the Lattice $\protect\cite{Ryan}$ (quenched (upper line)
and  unquenched (lower line)). }
\begin{center}
\begin{tabular}{|l|l|l|}
Meson & This model & Expt/Lattice \\
\hline 
$\pi^+$      & 131        & $130.7\pm 0.1\pm 0.36$    \\
\hline
$K^+$        & 161        & $159.8\pm 1.4\pm 0.44$        \\
\hline
$D^+$        & 211        & 203$\pm$ 14 \\
             &            & 226$\pm$ 15 \\
\hline 
$D^+_s$      & 222        &  230$\pm$ 14 \\
             &            &  250$\pm$ 30 \\
\hline   
$B^+$        & 180        &  173$\pm$ 23 \\
             &            &  198$\pm$ 30 \\
\hline   
$B^0_s$      & 196        &  200$\pm$ 20 \\
             &            &  230$\pm$ 30 \\
\hline
$B^+_c$      & 398        &         \\
\end{tabular}
\label{tab:fit}
\end{center}
\end{table}

\newpage


\begin{figure}[t]
\noindent Fig.1: Meson mass operator. 

\vspace*{1cm}
\noindent Fig.2: Hadronic decay $H_{13} \to H_{12} + H_{23}$. 

\vspace*{1cm}
\noindent Fig.3: Nonlocal coupling of meson, quark and photon fields: 

\noindent one-photon vertex (3a), two-photon vertex (3b). 

\vspace*{1cm}
\noindent Fig.4: Diagrams contributing to the electromagnetic 
form factor $F_{\pi}(Q^2)$ of the pion: 

\noindent triangle (4a) and bubble diagrams (4b and 4c). 

\vspace*{1cm}
\noindent Fig.5: Diagrams contributing to the processes 
$H\to \gamma^\ast\gamma^\ast$: 

\noindent triangle (5a), bubble (5b and 5c) 
and tadpole diagrams (5d). 

\vspace*{1cm}
\noindent Fig.6: Diagrams contributing to the decay amplitude 
$D(B)^+\rightarrow \sigma \pi^+$: 

\noindent direct (6a) and 
polar (resonance) contributions (6b).  

\vspace*{1cm}
\noindent Fig.7: Electromagnetic form factor $F_\pi(Q^2)$ of the pion 
in comparison to data taken from \cite{Volmer} (JLAB), 
\cite{Brauel} (DESY), \cite{Amendolia} (CERN) and to other 
theoretical calculations: QCD sum rules \cite{Nesterenko_Radyushkin}, 
light-cone quark model \cite{Gardarelli}, NJL model with a 
separable $q \bar q$ interaction \cite{Gross} and QCD modeling 
approach based on the solution of
the Dyson-Schwinger equations \cite{Maris}. 

\vspace*{1cm}
\noindent Fig.8: Results for the form factor 
$Q^2 F_{\pi\gamma\gamma^\ast}(Q^2)$ in comparison to
experimental data taken from \cite{Behrend} (CELLO),  
\cite{Gronberg} (CLEO) and to other theoretical calculations: 
hard scattering approach (HSA) \cite{Jakob}, 
QCD sum rules \cite{Radyushkin} and perturbative light-cone
QCD \cite{Cao}. 

\vspace*{1cm}
\noindent Fig.9: Our result for the form factor 
$Q^2 F_{\pi\gamma^\ast\gamma^\ast}(Q^2,\omega)$ 
for different values of the asymmetry parameter $\omega$. 

\vspace*{1cm}
\noindent Fig.10: Our result for the form factor 
$Q^2 F_{\pi\gamma\gamma^\ast}(Q^2)$. We also indicate the separate
contribution of the dominant triangle ($\Delta$) diagram.  

\vspace*{1cm}
\noindent Fig.11: Our result for the form factor 
$Q^2 F_{\sigma\gamma^\ast\gamma^\ast}(Q^2,\omega)$ 
for different values of the asymmetry parameter $\omega$.

\end{figure}

\newpage

\begin{figure}[t]
\vspace*{-6cm}
\centering{\
\epsfig{figure=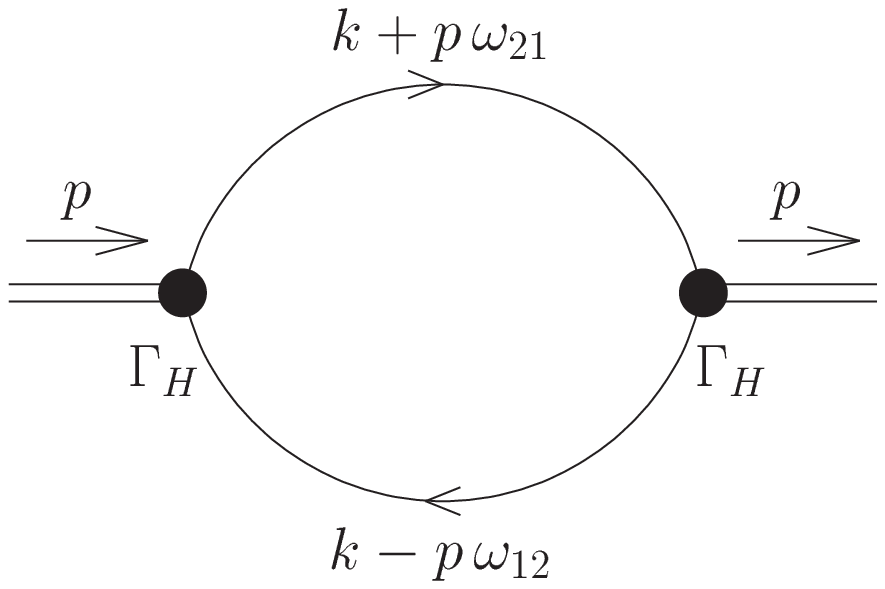,height=21cm}}

\vspace*{-9.5cm}

\centerline{\bf Fig.1}

\vspace*{-2cm}

\centering{\
\epsfig{figure=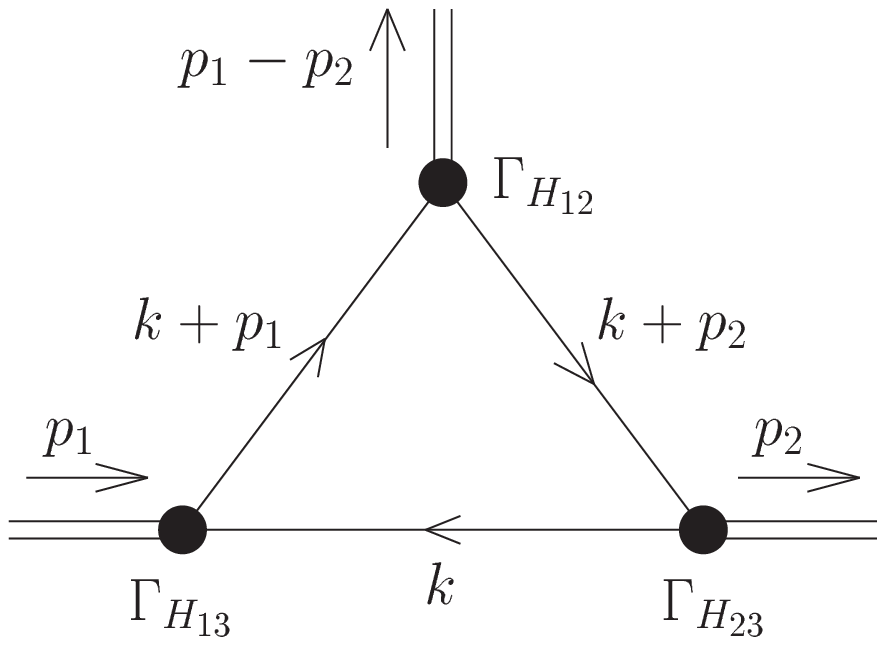,height=21cm}}

\vspace*{-11cm}

\centerline{\bf Fig.2}

\vspace*{-.5cm}

\centering{\
\epsfig{figure=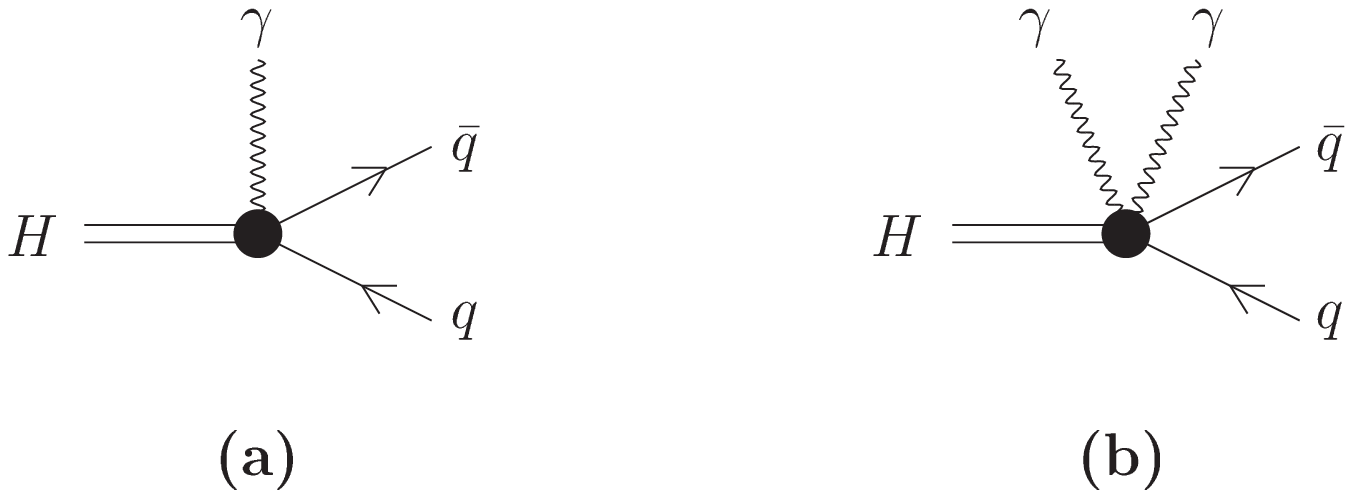,height=21cm}}

\vspace*{-14cm}

\centerline{\bf Fig.3} 
\end{figure}

\newpage

\begin{figure}
\centering{\
\epsfig{figure=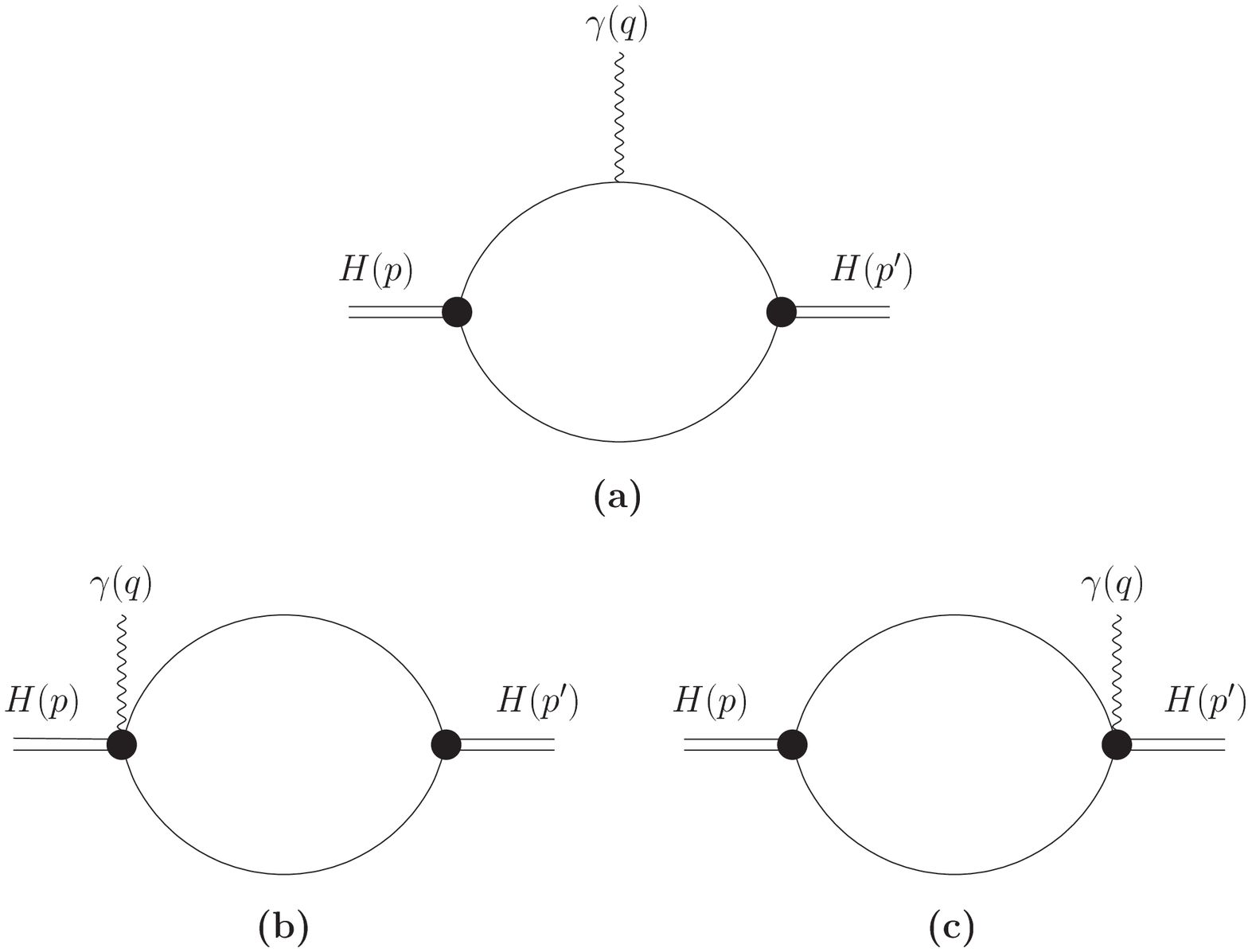,height=21cm}}
\end{figure}

\vspace*{-6cm}

\centerline{\bf Fig.4}

\newpage

\begin{figure}
\centering{\
\epsfig{figure=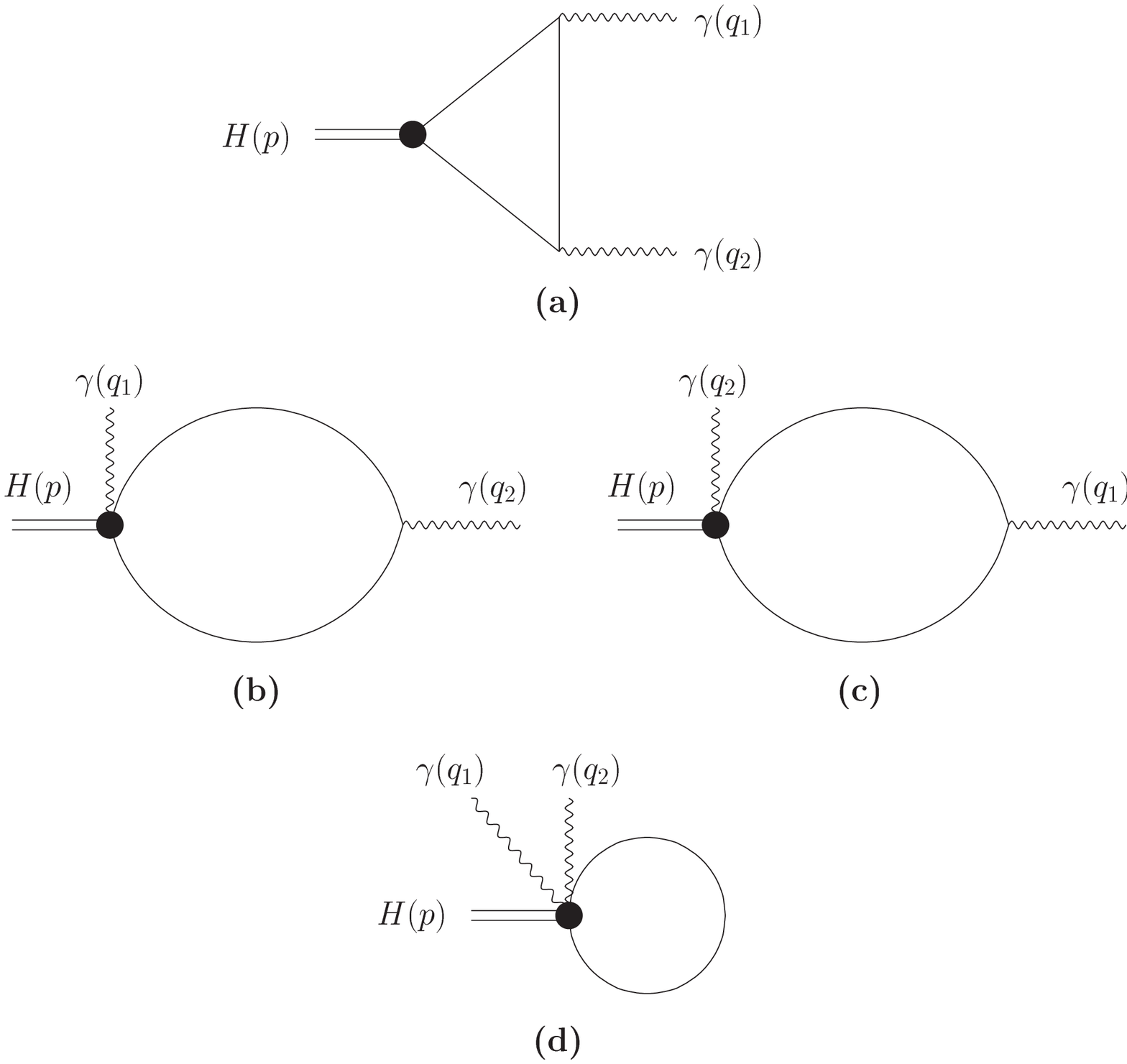,height=21cm}}
\end{figure}

\vspace*{-1cm}

\centerline{\bf Fig.5}

\newpage

\begin{figure}
\centering{\
\epsfig{figure=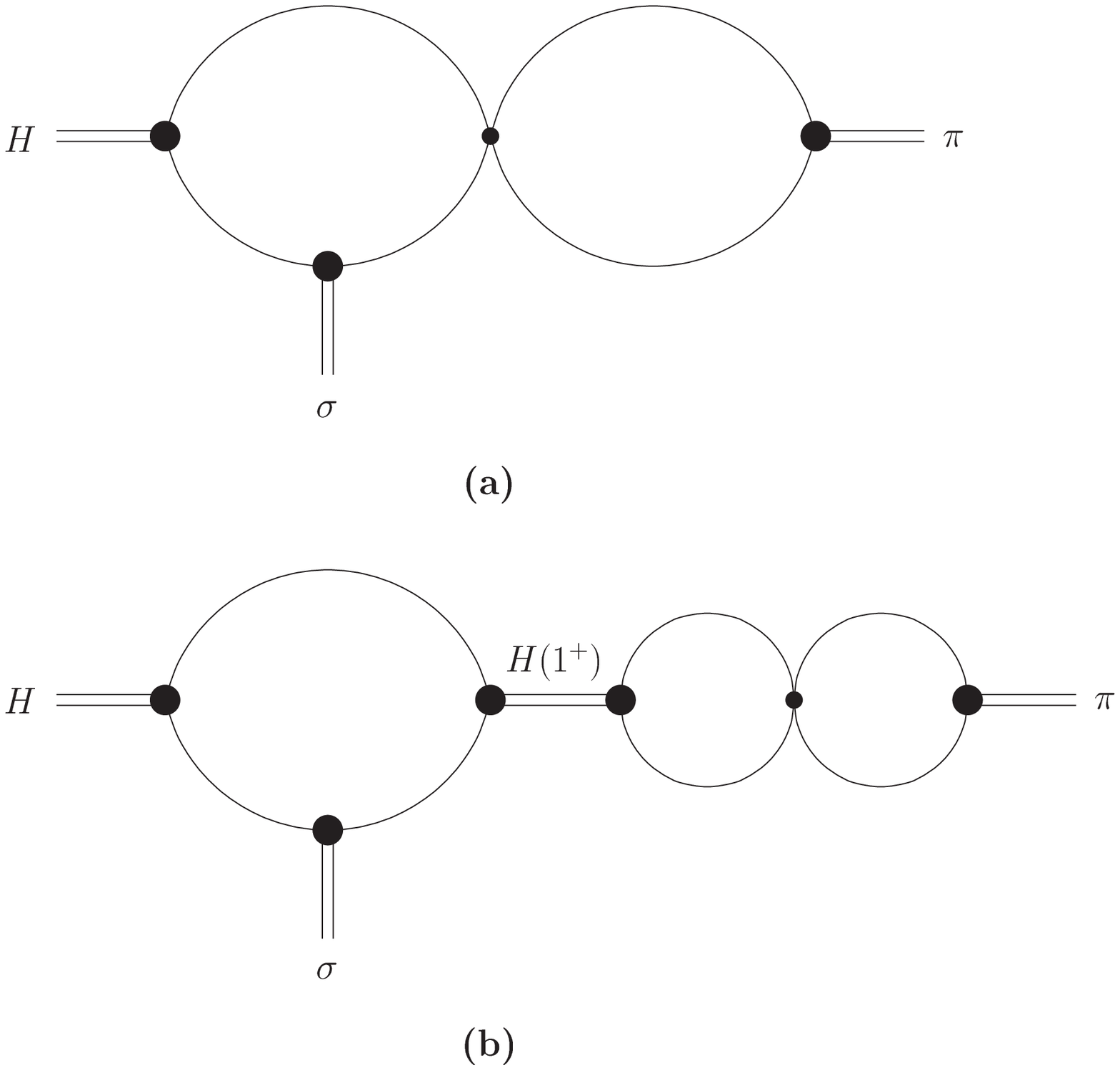,height=21cm}}
\end{figure}

\vspace*{-1cm}

\centerline{\bf Fig.6}

\newpage

\begin{figure}
\centering{\
\epsfig{figure=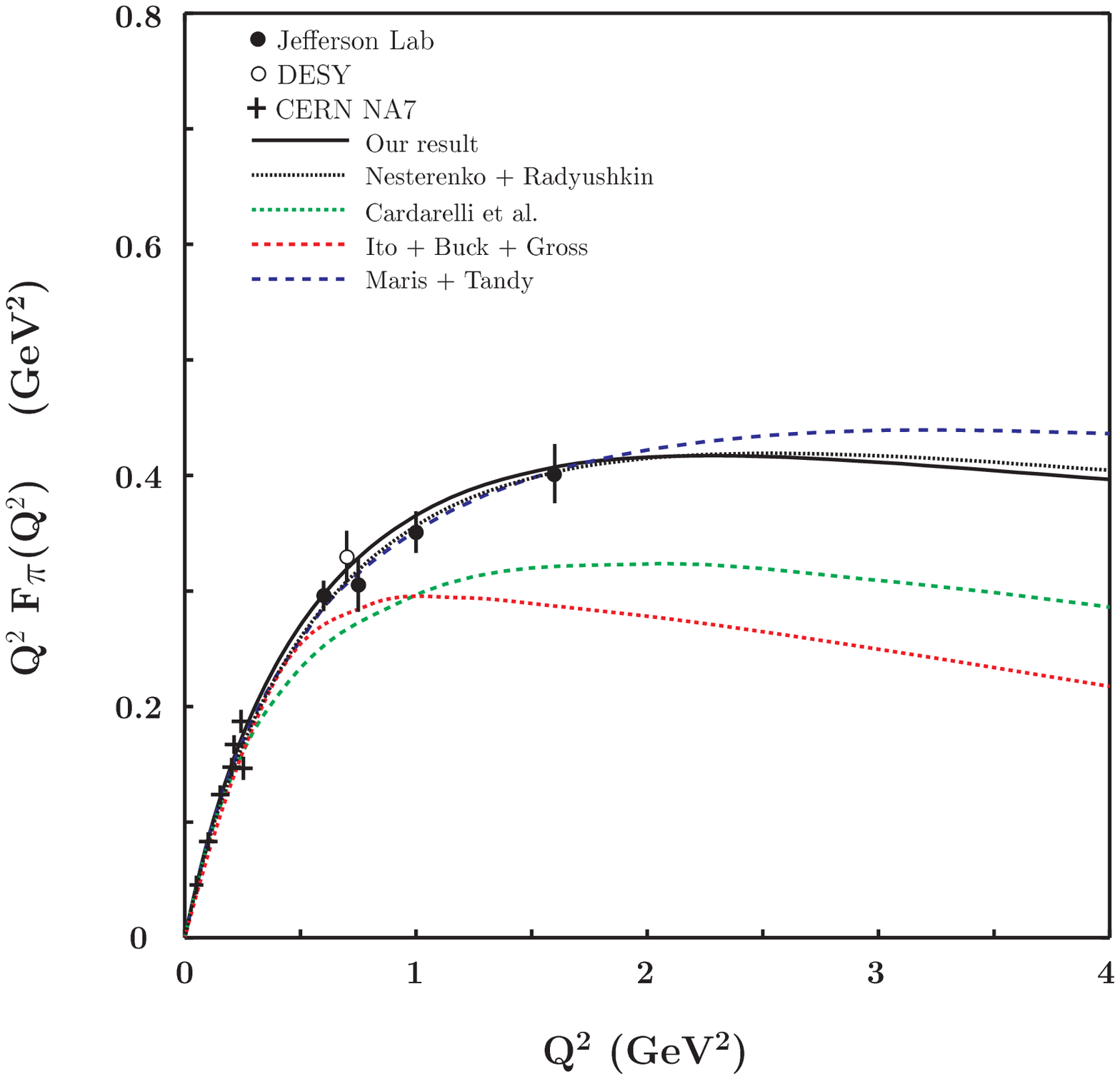,height=21cm}}
\end{figure}

\vspace*{-2cm}

\centerline{\bf Fig.7}

\newpage

\begin{figure}
\centering{\
\epsfig{figure=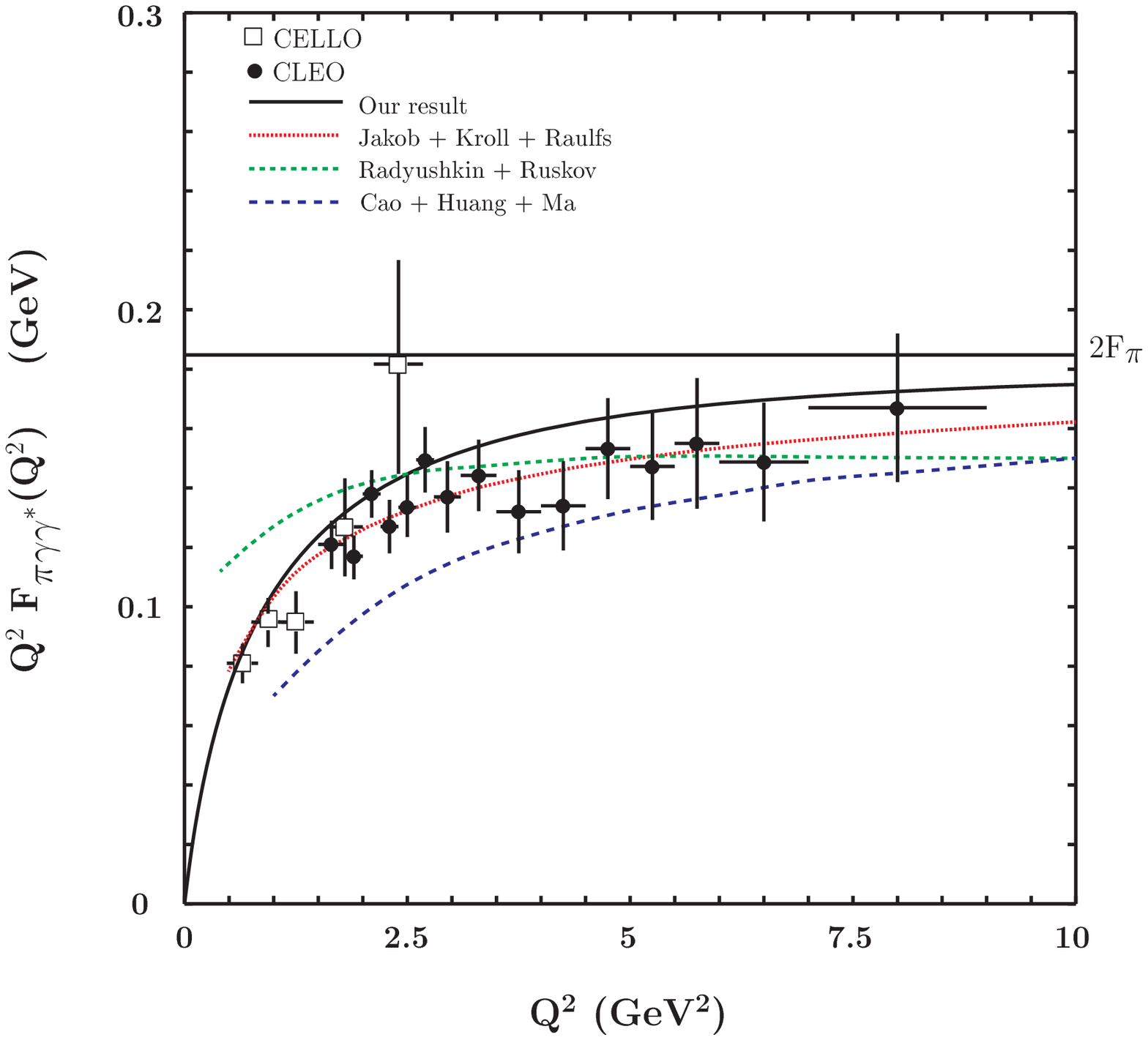,height=21cm}}
\end{figure}

\vspace*{-1cm}

\centerline{\bf Fig.8}

\newpage

\begin{figure}
\centering{\
\epsfig{figure=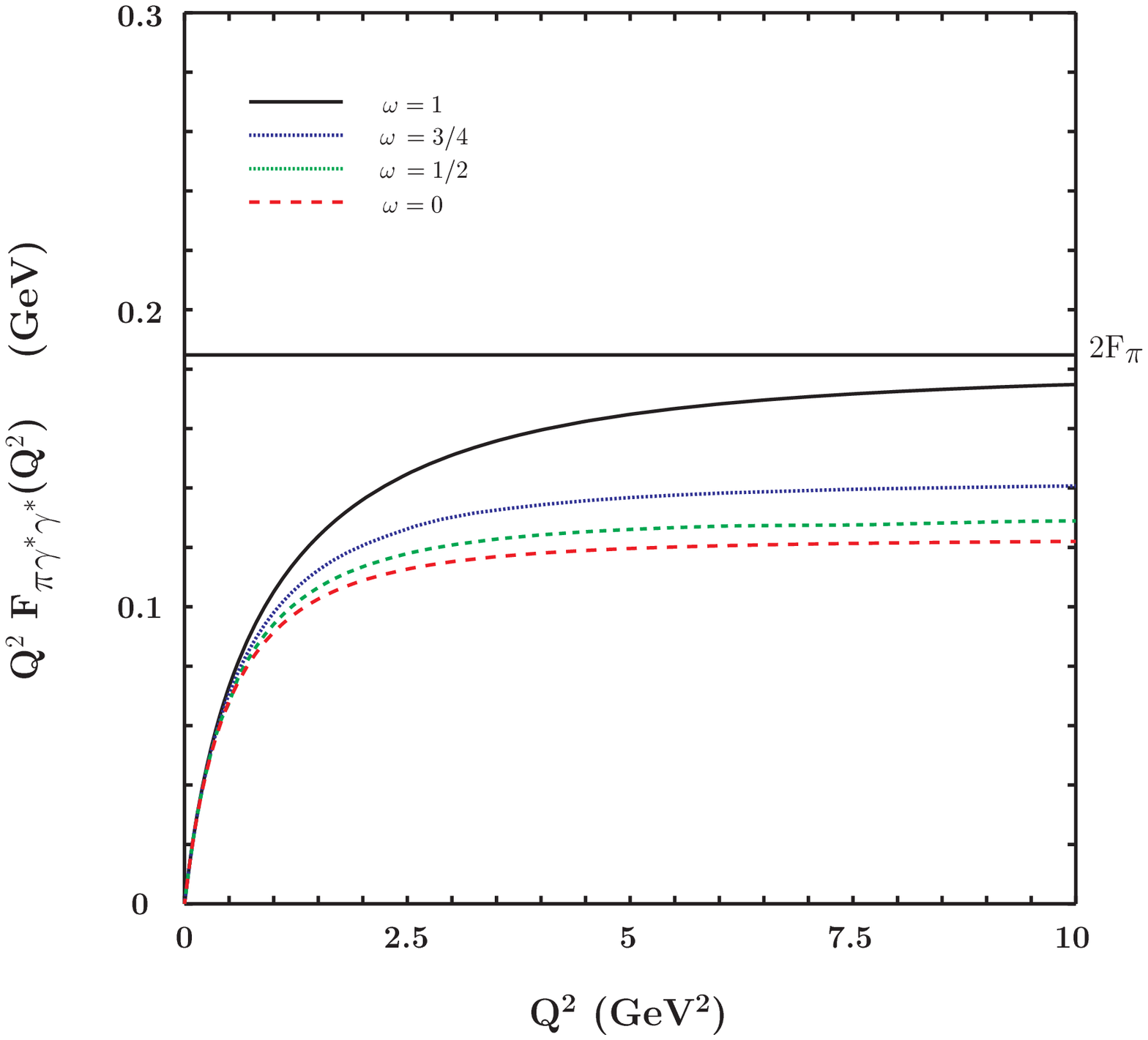,height=21cm}}
\end{figure}

\vspace*{-1cm}

\centerline{\bf Fig.9}

\newpage

\begin{figure}
\centering{\
\epsfig{figure=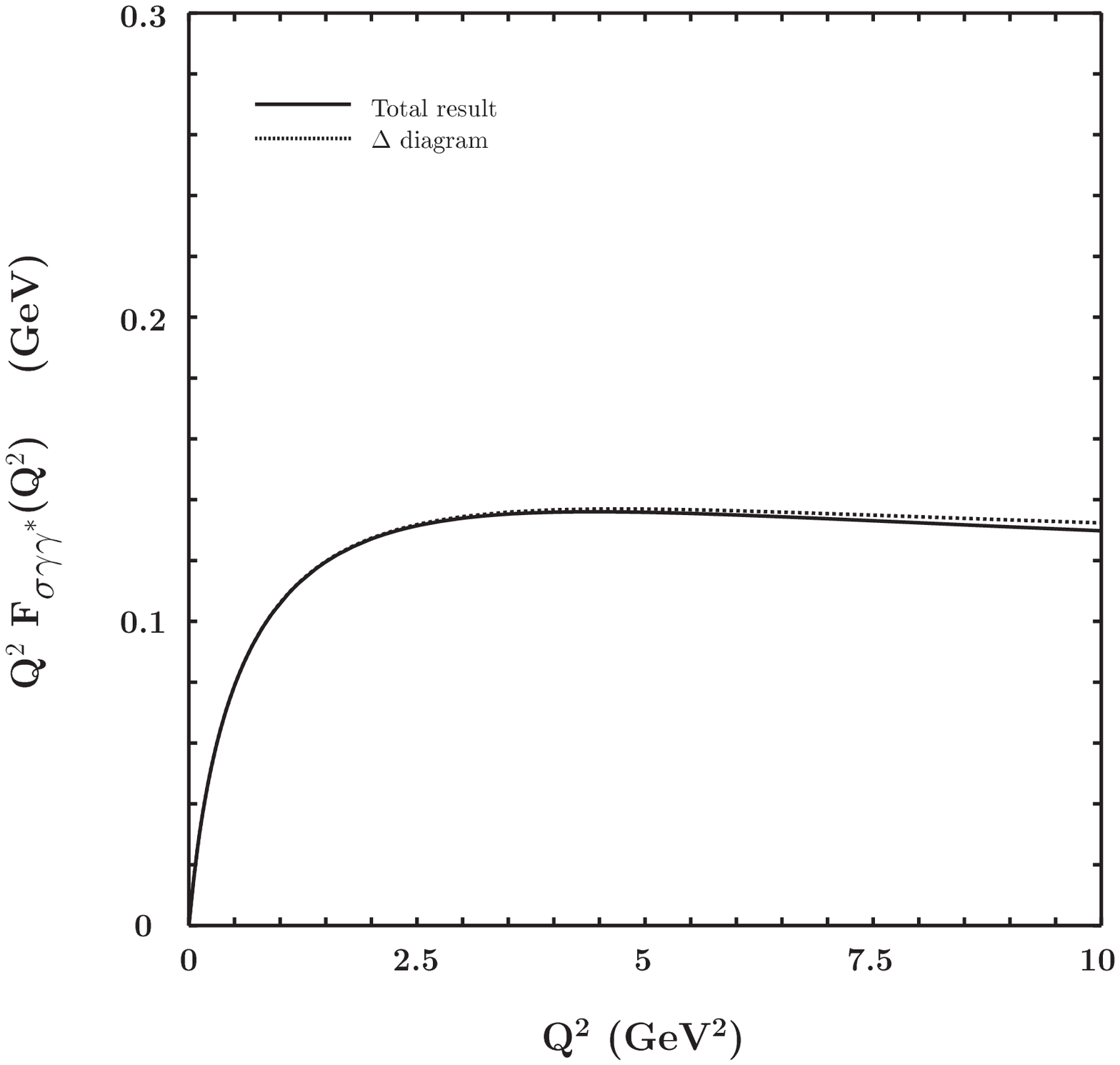,height=21cm}}
\end{figure}

\vspace*{-1cm}

\centerline{\bf Fig.10}

\newpage

\begin{figure}
\centering{\
\epsfig{figure=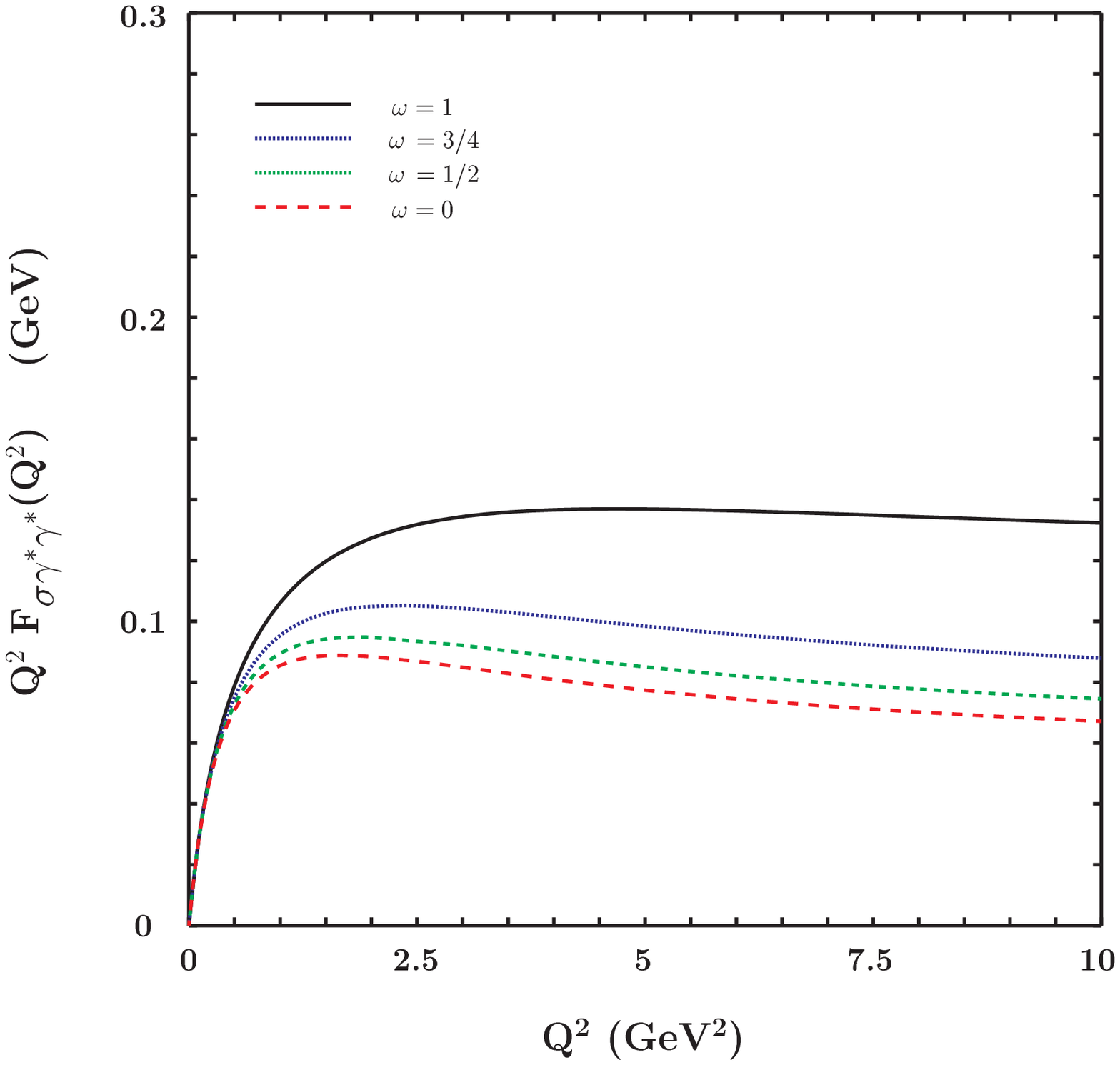,height=21cm}}
\end{figure}

\vspace*{-1cm}

\centerline{\bf Fig.11}

\end{document}